\documentclass[preprint]{aastex}

\slugcomment{Accepted for publication by AJ}

\shorttitle{Dust Opacities in Spiral Galaxies}
\shortauthors{Holwerda et al.}

\begin{document}

\title{The Opacity of Spiral Galaxy Disks IV:\\
Radial Extinction Profiles from Counts of Distant Galaxies seen through Foreground Disks.}

\author{B. W. Holwerda\altaffilmark{1,2}, R. A. Gonzalez \altaffilmark{3} and Ronald J. Allen\altaffilmark{1}}

\email{holwerda@stsci.edu}
\and
\author{P. C. van der Kruit \altaffilmark{2}}

\altaffiltext{1}{Space Telescope Science Institute, Baltimore, MD 21218}
\altaffiltext{2}{Kapteyn Institute, Groningen Landleven 12, 9747 AD Groningen, the Netherlands.}
\altaffiltext{3}{Centro de Radiastronom\'{\i}a y Astrof\'{\i}sica, Universidad Nacional Aut\'{o}noma de M\'{e}xico, 58190 Morelia, Michoac\'{a}n, Mexico}
 
\begin{abstract}
Dust extinction can be determined from the number of distant field galaxies seen through a spiral disk. 
To calibrate this number for the crowding and confusion introduced by the foreground image, 
\cite{Gonzalez98} and \cite{Holwerda04} developed the ``Synthetic Field Method'' (SFM), which 
analyses synthetic fields constructed by adding various deep exposures of unobstructed background 
fields to the candidate foreground galaxy field.

The advantage of the SFM is that it gives the average opacity for area of galaxy disk without assumptions 
about either the distribution of absorbers or of the disk starlight. However it is limited by low statistics of 
the surviving field galaxies, hence the need to combine a larger sample of fields. This paper presents 
the first results for a sample of 32 deep HST/WFPC2 archival fields of 29 spirals.
 
The radial profiles of average dust extinction in spiral galaxies based on calibrated counts of distant field 
galaxies is presented here, both for individual galaxies as well as for composites from our sample. 
The effects of inclination, spiral arms and Hubble type on the radial extinction profile are discussed. 

The dust opacity of the disk apparently arises from two distinct components; an optically thicker 
($A_I = 0.5-4 \ {\rm mag}$) but radially dependent component associated with the spiral arms and a relatively 
constant optically thinner disk ($A_I \approx 0.5 \ {\rm mag.}$). These results are completely in agreement 
with earlier work on occulted galaxies. The early type spiral disks in our sample show less extinction than 
the later types. LSB galaxies and possibly Sd's appear effectively transparent. The average color of the 
field galaxies seen through foreground disks does not appear to change with radius or opacity. This grey 
behaviour is most likely due to the patchy nature of opaque clouds. The average extinction of a radial 
annulus and its average surface brightness seem to correlate for the brighter regions. This leads to 
the conclusion that the brighter parts of the spiral disk, such as spiral arms, are also the ones with the 
most extinction associated with them.
\end{abstract}

       
\section{Introduction}

The optical depth of spiral disks has been the topic of many and 
varied studies since the claim by \cite{Holmberg58} that they 
are transparent. The subject became controversial when \cite{Disney89} and 
\cite{Valentijn90} argued that disks were virtually opaque. At the 
Cardiff meeting \citep{Cardiff94} many possible methods to 
attack the problem were proposed. 
The dust disks of spirals may obscure objects in the high redshift 
universe \citep{Alton01,Ostriker84} or conceal mass in their disks 
\citep{Valentijn90,Cuillandre01}. An excellent review of the current 
state of knowledge on the opacity of spiral disks is given by \cite{Calzetti01}.

Early approaches to this subject were presented in \cite{Cardiff94} and more recent developments are:\\
(1) disks are more opaque in the blue \citep{Tully98,Masters03}, 
(2) they are practically transparent in the near infrared \citep{Peletier92, 
Graham01b}, making these bands the best mass-to-luminosity estimator \citep{deJong96}, 
(3) disks are practically transparent in the outer parts but show 
significant absorption in the inner regions \citep{Valentijn94,Giovanelli94}, 
(4) the extinction correlates with galaxy luminosity \citep{Giovanelli95,Tully98,Masters03}, and 
(5) spiral arms are more opaque than the disk \citep{Beckman96,kw00a}
The majority of these studies are based on either inclination effects on the light distribution of a large 
sample of disks or a dust and light model to fit the observed profiles.
 
While there is some agreement on the view that spiral disks are substantially optically thick 
in their central regions and become optically thin in their outer parts, the 
exact radial extinction profile remains uncertain. 
Most measurements to date use the disk light itself to measure the extinction,  
and consequently require an assumption on the relative distribution of dust and light in a spiral disk.

The extinction in a disk can be derived from far infrared and sub-mm emission arising from 
the cold dust in disks. However, these methods make the assumption that the emission characterises 
the dust in the disk. But the far infrared and sub-mm emission is likely to be dominated by the warmest 
component of the dust, which tends to be the smaller grains on the outside of molecular clouds facing 
an energy source \citep{Dale02,Helou00}. In this case the far-infrared and sub-mm emission will 
underestimate the average opacity. \cite{Mayya97} mention this in their estimate of gas-to-dust ratio 
in spiral disks based on IRAS observations. 

To obtain a better characterisation of the absorption in a spiral disk without knowledge of the distribution 
of stars and dust in the disk, a known background source is needed. \cite{kw92} proposed to use an 
occulted galaxy for this purpose, assuming it has a symmetric light distribution \citep{kw99a,kw00a,kw00b,kw01a,kw01b}. 
\cite{Gonzalez98} and the companion paper of this work \citep{Holwerda04} use the number of distant 
galaxies in the field as their background source, calibrating this number with simulations following their 
``Synthetic Field Method'' (SFM). In \cite{Holwerda04}, we describe the details of this method 
\footnote{For the remainder of the paper {\it field galaxies} mean the distant background objects we count 
and {\it foreground galaxy} refers to the galaxy disk through which these distant galaxies are seen.}.

Both the ``Occulting Galaxy Method'' and the SFM have the benefit of not using the disk's own light to 
measure the extinction. The drawbacks of the Keel and White method are the assumption of symmetry 
of both galaxies and the few suitable pairs available. The SFM is more universally applicable but limited 
by the poor statistics. It does however not need to assume any distribution of either the disk's light or the 
absorbers in it. The SFM does require high-resolution images from HST and remains limited to the arms 
and disk of spiral galaxies due to crowding.  
 
In this paper we report radial extinction profiles for spiral disks of different Hubble types based on 32 
Hubble Space Telescope, Wide Field Planetary Camera 2 (HST/WFPC2) fields in 29 galaxies of 
Hubble types Sab and later. In section 2 we describe the sample and its selection in detail.
The SFM is briefly outlined in section 3; a more complete description of the method and the recent 
improvements we have made to it is given in the companion paper \citep{Holwerda04}. In section 4 we discuss 
the radial profile of average opacity for individual galaxies and, in section 5, 
the composite average radial extinction profile for our entire sample.
The effects of inclination, spiral arm prominence, and Hubble type on the profiles  
are discussed as well. Section 6 discusses the average (V-I) color of the field galaxies we found and 
section 7, the tentative relation between average surface brightness and opacity. We discuss some 
of the implications of our results, and end with the conclusions on disk opacity from the numbers and 
colors of field galaxies seen through them.

\section{HST Archive Sample}

Our sample of HST/WFPC2 fields was selected from the MAST archive at STSCI, based 
on criteria for both the target galaxy and the HST data. The total exposure times and original proposal ID and reference are listed in Table 1, the basic data on the galaxies in Table 2. The total solid angle of this sample is 146 arcmin$^2$.

The foreground galaxy should be a spiral, ideally face-on, spanning 
enough sky to cover a significant number of field galaxies. This solid angle 
constraint limits the maximum distance for application of the SFM to approximately 30 Mpc. 
HST starts to resolve the disk population of spirals at close distances, making 
the field too crowded for field galaxy identification (see \cite{Gonzalez03}). 
This imposes a minimum distance of a few Mpc. 
The galaxies are type Sab and later, plotted in Figure 1, according to the RC3 \citep{RC3}. 
No limit on inclination was imposed as long as spiral arms could
be discerned.

The majority of this sample is from the Distance Scale Key project.
The project's observing strategy \citep{KeyProjectObs,KeyProject}
was geared towards maximizing the number of Cepheid variables detected. This resulted in the selection of fields in the optical disk of face-on later-type spirals with a prominent arm visible. 
These selection criteria are to be considered when interpreting the opacity measurements of these fields.

The sample of spiral disks has multi-epoch imaging in 2 photometric band available, as the original science driver for many of these data sets was to sample the Cepheid light curves.
The reacquisition resulted in a slight shift in the pointing 
at each epoch. The unintentional 'dither' allowed us to drizzle the combined 
images to a pixel scale of 0.\arcsec05. The data reduction is described in 
detail by \cite{Holwerda04}. \cite{Gonzalez03} predicted that improved resolution would mainly benefit the statistics of distant galaxy counts in the Local Group. However, as the sample spans a range of distances and the statistics are likely to be poor, the maximum possible sampling was selected.
The PC chip was not used for analysis because it often lies on the most 
crowded region, its noise characteristics are different from the WF chips, 
and there are fewer reference fields available for the SFM.

This sample of HST pointings was selected as it is reasonably uniform, which 
allows similar processing using Hubble Deep Field (HDF) background fields.
The HST/WFPC2 pointing should neither be on the center of the galaxy where crowding
is too much of a problem, nor be outside the disk of the target
galaxy where the expected opacity is likely too low to be measured with the SFM.
We selected pointings with the F814W (I) and F555W (V) filters as
the identification of objects is based on several parameters characterising structure 
and on the V-I color.
A minimum exposure time of about 2000 seconds in both 
filters was adopted. The choice of filters and exposure time was based 
on our earlier experiences with spiral and irregular galaxies \citep{Gonzalez98, Holwerda02dunk}. 
and to maximize the number of suitable fields.

\section{SFM: Calibrating the field galaxy numbers.}

\cite{Holwerda04} describe in detail the data-reduction and the automated SFM, so we will give only a short summary here. To calibrate the numbers of field galaxies found in the science fields, simulated or ``synthetic'' fields are made. These are the original science fields with an extincted Hubble Deep Field added. The numbers of simulated field galaxies found suffer from the same confusion and crowding as the number from the science field and therefore depend solely on the dimming applied.
A series of simulations give us the relation between field galaxy numbers and applied dimming. The following relation is fitted to the simulated numbers:
 
\begin{equation}
A = -2.5 \ C\ log \left({N \over N_0 }\right) 
\end{equation}
 
\noindent $A$ is the dimming in magnitudes, $N$ the number of field galaxies in a simulation, $N_0$ and $C$ are the normalization and slope of the fit (i.e. the number of field galaxies expected with no extinction and how this number diminishes with increasing extinction). The slope C characterises the effects of crowding and confusion on the relation between opacity and galaxy numbers. It is usually slightly over unity but the unique character of each field led us to characterise C using the simulations in each field or combination of fields.
The intersection between this relation and the {\it real} number of field galaxies gives us the average extinction ($A_I$) typical for the solid angle where these field galaxies were found. The field galaxies are identified in the science fields, using automated selection, based on structural parameters and (V-I) color, together with a visual verification. 
In the synthetic fields, the visual check of objects was substituted by anti-correlating the automatically selected objects with the automatic selection in the science field, removing both the real distant galaxies as well as contaminants and leaving only added objects.

The uncertainty in the number of field galaxies from the science field is a combination of the Poisson uncertainty and the uncertainty due to field galaxy clustering. The uncertainty in the simulated numbers is Poisson only as they come from a known background, the Hubble Deep Fields. The uncertainty in average opacity is derived from the uncertainties in field galaxy numbers based on the Poisson uncertainty as expressed by \cite{Gehrels85}, and the clustering uncertainty based on the 2-p correlation function found by \cite{Cabanac00} and equation 1. The crowding and confusion bias is calibrated with the simulations. For a detailed error discussion of the SFM see \cite{Holwerda04}, \S 4.3.

\subsection{Galactic Extinction}

A difference in dust extinction from our own Galaxy between the reference fields (HDF-N/S) and the pointing at the foreground galaxy will introduce a bias in the extinction measurement of the specific disk. \cite{Gonzalez99} used galaxy counts to measure the extinction towards GRB 970228. and found excellent agreement with other measurements of Galactic extinction \citep{Burstein78b,Schlegel98}. 
\cite{Schlegel98} produced an all-sky map of Galactic extinction based on COBE and IRAS maps and we use their values for Galactic extinction (Table 2). 
Most galaxies in the sample do not show a significant difference in Galactic extinction compared to the average of the Galactic extinction towards the HDF-N/S ($A_I$ = 0.039 mag). However, the numbers of galaxies from each science field were nevertheless corrected for the difference in Galactic extinction using equation 1. 

\section{Radial opacity measurements in individual WFPC2 fields}

\cite{Gonzalez98} presented results based on the SFM for individual Wide Field chips, characterised as ``arm'' or ``disk'' regions based on predominance in the chip. \cite{Holwerda04} segmented the WFPC2 mosaics based on morphological component (arm, inter-arm, disk) or projected radius from the center. However, the statistics from individual WFPC2 fields barely allow any meaningful opacity measurements for solid angles smaller than a single WF field. In Table 3 we present opacity values for the projected radial annuli in each foreground galaxy in our sample. The radii are expressed in $R_{25}$, half the $D_{25}$ from RC3 catalog \citep{RC3}. The error-bars are computed from the uncertainties in the numbers of real and simulated field galaxies from counting and clustering \citep{Holwerda04}. For the galaxies for which we have two WFPC2 pointings (NGC5194, NGC3621 and NGC4414), the radial extinction from the combined counts is also shown. 

\section{Average Radial Opacity Plots}

Estimates of extinction in a galaxy disk based on a single WFPC2 chip suffer from poor statistics \citep{Gonzalez98, Holwerda04}, which make a radial dependence hard to establish (See Table 3 and \cite{Holwerda04}). This led us to apply the SFM to a large sample of foreground galaxies in order to estimate the general extinction properties of galaxy disks from the combined numbers of field galaxies seen through these disks. We caution against averaging the values in Table 3 to derive average profiles.
Figure 2 shows the radial opacity plot from all our fields combined. The numbers of field galaxies from both the science fields and the simulations from all fields were combined based on their projected radial distance from the respective galaxy's centers,\footnote{The values used to deproject the distances on the sky to radial distance to the galaxy's center are presented in Table 2.} expressed in $R_{25}$. The top panel in Figure 2 shows the total combined solid angle for each radial bin for which the average opacity was determined. The middle panel shows the number of field galaxies from the science field and the average number found in the simulations without any dimming (A=0). The bottom panel shows the opacity for each bin derived from the intersection of equation 1 fitted to the simulations and the real number of galaxies. 

The solid angle or the number of simulated galaxies without any dimming is a good indicator for the reliability of our opacity estimate, as reflected by the error bars in the bottom panel of Figure 2.
The estimates are limited at small radii by the high surface brightness and crowding of the foreground galaxy center, effectively masking some of the solid angle available at those radii.
At higher radii the uncertainty comes from a lack of solid angle covered (See Figure 2, top panel). This is a selection effect of our sample, as most of the WFPC2  fields were pointed at the optical disk of the galaxy. 

In order to determine the effects on average radial opacity of disk inclination, prominence of spiral arms or Hubble type, radial opacity plots of subsets of our sample were constructed. The solid angle used is then some fraction of those in the top panel in Figure 2, which consequently increases the uncertainty in the opacity measurement. To counter this, instead of the radial bins of 0.1 $R_{25}$ (Figure 2), radial binning of 0.2 $R_{25}$ was applied. 

\subsection{Inclination effects}

The inclination of the foreground disks affects the measured opacity, but the amount of this effect depends on the dust geometry. In the case of a uniform thick screen, the path length attenuating the field galaxies is increased. However in the case of a screen of dark clouds in the disk, the effect is on the apparent filling factor of clouds. Because the correction depends on which dust geometry is assumed, we present the radial results (Figure 2) without any correction and explore corrections using several different models, illustrated in Figure 3.
The homogeneous screen results in a multiplicative factor cos(i) to be applied to the opacity value (A) 
or in a correction to the number of field galaxies found in the science field, as follows: 

\begin{equation}
N_{\perp} = {N_i^{cos i} \over N_0^{cos i -1}}
\end{equation}

\noindent The other  models consider a screen with fully opaque clouds or patches. Depending on the thickness of these clouds, the apparent filling factor depends differently on inclination. In these models, the clouds have an average oblateness $\epsilon$ ($\epsilon = 1-{b \over a}$) with major axis a and minor axis b of the clouds. Following the geometry of Figure 3, the relation between the number of field galaxies, seen through the foreground disk face-on ($N_{\perp}$), at an inclination i ($N_i$), and the average number of field galaxies in the field behind the foreground galaxy ($N_0$) can then be expressed as:

\begin{equation}
N_{\perp} = [ \ 1\ - \ \epsilon] \ [1 \ - \ cos(i)\ ] \ N_0 + \left\{ \ \epsilon \ [ 1-cos(i) ] \ + \ cos(i) \right\} \ N_i 
\end{equation}

\noindent where the extreme cases for $\epsilon$ are spherical clouds ($\epsilon=0$) and flat patches ($\epsilon=1$). 
The oblateness $\epsilon$ parameterises the ratio between the scale height of the dust and its extent in the plane of 
the disk. In images of edge-on disks, the visible dust lanes are confined to a thinner disk than the stars. It is therefore 
likely that $\epsilon$ is not 0. For purposes of illustration we use a value of 0.5 in Figure 8.

Figure 4 shows the radial profile from Figure 2 corrected for both extreme cases and uniform screens in all galaxies. The contributions from each galaxy to the composite radial profile were corrected according to equation 2 and equation 3 before addition. Figure 5 shows the total uncorrected radial profile for four subdivisions of our sample based on inclination. From Figure 5 it seems clear that other effects are much more important than the inclination of the foreground disk. For this reason we ignoring the effects of inclination on our measurements.
As the effects of inclination are debatable on the basis of preferred dust geometry, we present further results without inclination corrections except where noted.

\subsection{The effect of spiral arms}

Another effect on the average radial opacity profile in Figure 2 is due to the presence of spiral arms. As our sample is predominantly from the Cepheid Distance Project, the WFPC2 images all feature spiral arms. If these are more opaque than the disk proper \citep{Beckman96,kw00a}, then the radial profile presented in Figure 2 could be biased towards higher opacities. Separate radial plots for the arm, the inter-arm part of the disk and outside any spiral arm part of the disk are shown in Figure 7 based on the counts from the typical regions in the entire sample, with the exception of the LSB galaxies UGC 2302 and UGC 6614. They were left out, as not much spiral structure can be discerned; they appear completely transparent (See Figure 8). 

\subsubsection{Segmenting images}

In order to differentiate between the effects of spiral arms and disks in the opacity plots, the images were segmented into crowded, arm, disk (inter-arm) and disk (outside arm). 
These regions were flagged in the mosaiced WFPC2 fields using the GIPSY (Groningen Image Processing System) function {\it blot}, in the same way as NGC1365 in \cite{Holwerda04}. The choice of typical regions was made in order to compare the arm and inter-arm results of \cite{kw00a} and \cite{kw00b}. A typical mask is presented in Figure 6 and masks for all galaxies in the sample are presented in \cite{mythesis}. It should be noted that this segmentation into typical regions is subjective, and based on those sections of the foreground galaxies covered by the WFPC2 observations we use.

\subsubsection{Radial extinction for typical disk regions}

Radial extinction profiles of the typical regions are presented in Figure 7.
The arm regions show much more opacity and a much more pronounced radial dependence of that opacity. There is a radial dependence as well for the inter-arm parts of the disk, although not as steep. The outside parts of the disk of the spiral galaxy, however, show little or no relation between opacity and radius. The opaque components of a spiral disk appear to be the spiral arms while the disk itself is more transparent but much more extended.
Figure 7 also shows the total solid angle of the radial annuli over which the opacities are determined. From these we can conclude that the regions deemed `arm' are not dominating the whole of the fields. 
The radial effect of the arms only becomes visible in the total radial opacity plot (Figure 2 and top left panel Figure 7), at the lower radii where the arms and the inter-arm region dominate.
 In Figure 8 we present the profiles corrected for inclination assuming an $\epsilon$ of 0.5. The general trends remain but now for somewhat lower opacity values. 

\subsubsection{Comparing to the ``Occulting Galaxy Method''}

\cite{kw00a} and \cite{kw00b} presented their extinction values from occulting galaxy pairs as a function of the radius, scaled with $R_{25}$. \cite{kw00a} compared ground-based photometry, and \cite{kw00b} used spectroscopic measurements of the occulted galaxy light. These extinction points are plotted in figure 9 for arm and inter-arm regions. The extinction curves from Figure 7 and the inclination corrected one from Figure 8 are plotted as well. Both the arm and the inter-arm extinction values as a function of radius agree well with the values obtained from the occulting galaxy technique. It is remarkable how well the results compare, considering they were obtained from completely different samples of spiral galaxies and using different techniques.
The values from the occulting galaxy technique are slightly lower than ours. There are several possible reasons for this. The spectroscopic results from \cite{kw00b} (triangles in Figure 9) possibly favour the more transparent regions in a disk (Domingue 2004, private communication).
But more importantly, the sample of occulting foreground galaxies consists of a different makeup of spiral galaxy subtypes than that of this paper. It should be noted that there are no galaxies common to both our sample and that of the occulting galaxy technique.
single galaxy in both our sample and the occulting galaxy method's. \cite{kw00b} noted that their later types (Sbc) seem more opaque, as do we (Figure 11). Figure 10 compares the results for the most prevalent Hubble subtypes in the occulting galaxy method (Sb and Sbc) with our results for those. The arm values seem to match up but there is a difference in the inter-arm results. It is unclear to us whether this points to a structural effect in either technique. A likely explanation is that the definition of ``inter-arm'' applies to slightly different regions in the spiral disks for \cite{kw00a,kw00b}, and his paper. This paper's definition of typical regions is based on the high-resolution mosaic, whereas the \cite{kw00a} is on their ground-based imaging. There is therefore the possibility that we include sections in the interarm regions which the occulting techniques would not resolve as ``inter-arm'', raising our values of opacity for those regions with respect to the occulting galaxy technique.

The SFM provides an independent verification of the occulting galaxy technique using a fundamentally different approach. In addition, a component was added to the distinction between arm and inter-arm parts of the disk, the ``outside'', meaning not directly enclosed by spiral arms. The fact that this component is not fully transparent raises the possibility of a dust disk extending beyond the spiral arms (Figure 8).

\subsection{Hubble Type}

With the effect of spiral arms on the opacity profile, the Hubble type is likely to be of influence on the profile. Figure 10 shows the extinction profiles averaged per Hubble type. Hubble types Sab through Sd are presented, as well as the average opacity profile of the two LSB galaxies in our sample. 
Hubble types Sab through Scd in our sample show disk extinction up to the $R_{25}$. The Sab result is tentative, owing to the poor statistics from only 2 WFPC2 fields.  
The Sd and LSB galaxies however appear effectively transparent. LSB galaxies were in fact assumed to be transparent by \cite{ONeil00} when discussing the morphology of field galaxies seen though them. However both profiles, Sd and LSB, are based on only two WFPC2 fields, accounting for the higher uncertainties. Comparing early type galaxies with later types, it appears that the later type galaxies (Sbc-Sc) show more extinction and at larger radii. 
The Sb galaxies show a bump which appears to be associated with higher extinction from the more tightly wound spiral arms.

Figures 12 and 13 show the radial profiles for arm and disk -both inter-arm and outside regions- for our sample divided into early and late spirals. A finer separation in Hubble type would have resulted in even higher uncertainties, due to lack of solid angle and hence statistics. The solid angle of each radial annulus is also plotted. Purely arm regions do not dominate the opacity profiles. However, the interarm regions are similar in behaviour (Figures 7 and 8). The inner parts of the profile are therefore more arm-like, while the remainder is disk dominated.

The radial dependence of arm and inter-arm regions is more pronounced in the case of earlier galaxies than for later types. The opacity in the spiral arms is substantially higher then in the disk, for both early and late types. The total profiles show bumps at 0.9 and 1.1 $R_{25}$ for the early and late types, respectively. In the case of the early types this seems due to the spiral arm contribution at that radius. The bump for the late types is not as significant, but might be related to the general position of spiral arms as well.

\section{Average color of the field galaxies}

The average color of the field galaxies found in the science fields can, in principle, tell us something about the actual dust geometry responsible for the drop in numbers. If there is a correlation between the average reddening of the field galaxies and the average opacity of the foreground spiral, then the dust extinction responsible for the drop in number of field galaxies is, at least in part, in the form of a diffuse screen, reddening the visible galaxies. If, however, the average color of the field galaxies does not change with opacity, then the drop in their number is likely due to fully opaque clouds with transparent sections between them to allow for the detection of the unreddened surviving field galaxies. \\

Our method influences the average color of the detected field galaxies in several ways. 
The V band (F555W) for the synthetic field background was constructed from the original HDF images in the F606W and F450W filters. However, \cite{Gonzalez98} estimated that this introduces a negligible error.
Crowding introduces blended objects in the synthetic fields. The synthetic counts are corrected for this effect but it will influence the average color. The simulated dust extinction we used was grey, so no preferred reddening was introduced in the synthetic fields. 
However, our automated selection procedure for field galaxy candidates selects against very blue objects, introducing a preference for red galaxies.
A similar selection effect may take place in the visual check of the science fields, as blue objects are treated as suspected foreground objects. 
Overall, these selection biases will cause the average color of the field galaxies in the science and synthetic fields to be redder than the average for an unobstructed field of distant galaxies. The effects of blended objects on the average colors of science and synthetic fields are not identical, as not all blends have been removed from the synthetic counts. Nonetheless, we can compare the trends of both these average colors with radius.


The average (V-I) color of the field galaxies found in the science fields, the simulation without extinction (A = 0), and the opacity derived from the galaxy numbers are plotted as a function of radius in Figure 14. The average (V-I) color of the field galaxies in the science fields does not change with radius and hence average opacity. The average color of the synthetic field objects appears to become bluer with radius. 
However, beyond 1.4 $R_{25}$ the number of objects is very small and the averages should be treated with caution. Comparing the average color of distant galaxies from science and synthetic fields for the inner part of the disk, the average of the science field objects is redder than the synthetic field average. 
This difference in average color is likely the result of blends with blue foreground objects being inadvertently included in the the selection of synthetic field objects. The number of objects from the synthetic fields was corrected for this effect (see also \cite{Holwerda04a} but the average color was not.
The science field objects do not suffer from this as they were checked visually for blended objects.
The field galaxies seen in the science fields are likely visible in parts of the disk which are nearly transparent 
or, alternatively, the dust screen in the disk behaves according to the ``grey'' extinction law. This is remarkable, considering the bias towards redder objects throughout our method.

Figure 15 shows the average color-extinction measurements for the field galaxies in the science fields. The Galactic extinction law is shown for comparison. Each point has been determined in a radial annulus, for all the fields combined (Figure 2) and for the typical regions (Figure 7, but with the finer radial sampling of Figure 2.). 
Without distinction between regions, no trend with opacity can be discerned for the average color.

The average color in the arm regions does not seem to rise much with opacity. The arm extinction is decidedly grey, something also found by \cite{Gonzalez98}. For the disk regions, no distinct trend can be seen, and they do not seem to follow the Galactic law very well. \cite{Gonzalez98} found the disk region of NGC4536 to be more Galactic in its reddening law. There is an average reddening with respect to the average (V-I) color of the HDF objects identified as galaxies by our algorithm without a foreground disk. This reddening is likely to be the result of the effects of the visual check and contamination of the color measurement by stray disk light.

These color-extinction relations for both arms and disk seem to be greyer than the Galactic extinction slope, favouring the possibility that at least part of the extinction is in opaque clouds. 
Uncertainties are such, however, that even for our increased statistics no good relation between opacity and color can be found.  

\section{Surface brightness and opacity}

The counts of distant galaxies were added per radial annulus, expressed in $R_{25}$. Additionally, the flux and solid angle from each field can be added for each radial annulus. The averaged surface brightness ($SB_I$) and the average opacity (A) from each radial bin in Figure 2, are plotted in Figure 16. These values for the radial bins per typical region in Figure 7 are plotted as well, but with finer sampling. There is a hint of a relationship between luminosity and extinction. This is consistent with
the relation found by \cite{Giovanelli95} and \cite{Masters03} between overall disk opacity and the galaxy's total luminosity.
The solid angles over which the surface brightness and opacity were averaged were selected by radius and not luminosity. Areas with different surface brightnesses are therefore combined at each radius, smoothing out any relation between opacity and surface brightness. Hence, most of the points in Figure 15 are around the same opacity and the range of surface brightness values is small. Nevertheless, the values of high extinction and brightness ($SB_I < 18.5$) do show some relationship between surface brightness and opacity, with arm regions displaying more extinction at the same surface brightness levels than the inter-arm disk. The disk values appear to show no correlation, but these values are per definition not for high surface brightness levels. 
And yet, together with the higher values of opacity found for the spiral arms, the points are not inconsistent with a relation between the opacity and luminosity. 
Opacity measurements in partitions of the WFPC2 images based on average surface brightness instead of radius should reveal any relation more clearly. This comparison between extinction and emission will be presented, in more detail, in a later paper.

\section{Discussion}

From the number of field galaxies found through the disks of spiral galaxies, a quantative picture of extinction as a function of radius can be found. The SFM is too limited by statistics to obtain a good result for individual fields. However a meaningful measurement can be derived from a combination of several galaxies. The effect of spiral arms on the radial extinction is quite distinct and dependent on Hubble type. 
From these radial plots it becomes clear that the dust in the disk is not one smooth layer sandwiched between the disk's stars but two separate components: radially dependent spiral arms and a more transparent but also extended disk. In addition, there is a highly opaque central bulge component.

Whether the average opacity is a result of dust clouds or a smooth screen of grey dust, is difficult to determine from the numbers, colors and luminosities of the field galaxies found through the disks of our sample. However, the relative independence on inclination of the average opacity measured from numbers of spiral galaxies and the grey relation between opacity and average field galaxy color, both point towards a patchy distribution of the absorbing dust in the disk. 
 
Assuming that the optical depth of the disk can be expressed as ${\rm \tau =  ln(1-f)}$, where f is the area filling factor, the average opacity of $A_I \approx 0.5$ would require a filling factor of 40\% of the disk, a figure that rises to 85 \% in the spiral arms. While the visible dust lanes can account for at least a part of these clouds, other dust clouds are likely embedded in the disks and arms. This patchy coverage explains the occasional distant galaxy seen through a spiral disk \citep{Roennback97, Jablonka98}, which has sometimes been used as anecdotal evidence for transparent disks. The disk is in fact relatively transparent where background galaxies are seen. However, by calibrating the number of the distant galaxies found, a very different picture emerges.

While this result is consistent with earlier findings, these values are likely upper limits, not lower ones. Any inclination correction will lower the face-on value for extinction. 
However, the patchy nature of dust extinction is likely responsible for a high variety of extinction values in the disk. Integrated measurements such as light profiles, however, will be affected according to the presented opacity values. A possible relation between extinction and brightness would also cast doubt on the fixed mass-to-light ratios generally assumed when modelling the kinematics of a spiral disk. 


\section{Conclusions}

The effects of dust extinction on the number of field galaxies found in our fields lead us to the 
following conclusions:\\
(1) The SFM gives an unbiased but uncertain measure of opacity for single WFPC2 fields (Table 3).\\
(2) On average, the disk of a spiral galaxy has an opacity in $I$ of $\approx$1 magnitude (Figure 2).\\
(3) The extinction measured from the number of field galaxies seems to be independent of 
the inclination of the foreground disk over the range in $i$ covered by our sample 
($10^{\circ} < i < 70^{\circ}$) (Figure 5). This is consistent with fully opaque flattened clouds 
covering a fraction of the area as the cause of the observed average opacity.\\
(4) The absorption profile for typical regions in the disk is strongly influenced by the spiral 
arms. Spiral arm regions are the most opaque and display a radial dependence. The disk 
regions enclosed by a spiral arm are also more opaque than other disk regions and display 
a similar radial dependence (Figure 7). \\
(5) The radial extinction curves derived from numbers of background galaxies and those 
reported by the occulting galaxy technique agree reasonably well (Figure 9 and 10), although 
a systematic effect in either (or both) technique seems to be present in the interarm results (Figure 10).\\
(6) Sc galaxies show much more opacity in their central regions than other types (Figure 11).\\
(7) All Hubble types earlier than Scd in our sample show substantial disk extinction in $I$ up 
to $R_{25}$ (Figure 11).\\
(8) The numbers from Sd and LSB galaxies are consistent with a transparent disk, but these 
measurements are limited by statistics. (Figure 11).\\
(9) Both early and late type spirals exhibit the extinction profiles of the two distinct components: 
a radially dependent one (arm and inter-arm), and a more extended disk (outside arms) (Figures 12 and 13).\\
(10) The average color of the distant galaxies identified in the science fields does not change 
with either radius or the opacity derived from their numbers (Figure 14).\\
(11) The grey nature of the absorption derived from field galaxy counts and colors holds for all 
typical regions (Figure 15). The absence of inclination effects and the grey nature of the absorption 
is consistent with strongly absorbing dense clouds masking off the distant galaxies. Their covering 
factor in the disk would be around 40\%, regardless of cloud sizes (which our technique cannot provide).\\
(12) The average surface brightness in radial annuli and the corresponding average opacity 
derived from distant galaxy counts appear correlated. They are consistent with a rise of opacity 
with surface brightness. (Figure 16). Although the range of surface brightness is small and the 
uncertainty in opacity rises with higher surface brightness, this is in good agreement with earlier 
results for bright galaxies. This would constitute a relation between dust mass and light.\\

The ``Synthetic Field Method'' has proven itself to be a useful, model-independent, technique to 
measure the total opacity of spiral disks. It can be applied to any spiral disk at intermediate 
distance for which there is high resolution imaging available. 
We will present further results for opacity as a function of surface brightness in a future paper.

\acknowledgments

The authors would like to thank the anonymous referee for his or her comments and 
Harry Ferguson for insightful discussions on the ``Synthetic Field Method''.
This research has made use of the NASA/IPAC Extragalactic Database 
(NED) which is operated by the JPL, CalTech, under contract with NASA.
This work is primarily based on observations made with the NASA/ESA 
Hubble Space Science Telescope (STScI) and obtained from the data archive at the 
Space Telescope Institute. Support for this work was provided by NASA 
through grant number HST-AR-08360 from STScI. 
STScI is operated by the association of Universities for 
Research in Astronomy, Inc., under the NASA contract  NAS 5-26555.
We are also grateful for the financial support of the STScI Director's 
Discretionary Fund (grant numbers 82206 and 82304), and of the Kapteyn 
Institute of Groningen University




\begin{deluxetable}{l l l l l l}
\tablewidth{0pt}
\tablecaption{HST Archive Data sample}
\tablehead{
\colhead{Name} 	& \colhead{Exp Time} 		& 					& & \colhead{Prop. ID.} 	& \colhead{Reference} \\ 
		& \colhead{$V_{F555W}$}	& \colhead{$I_{F814W}$} & & 					& 	}
\startdata
NGC 925 		& 26400.0		& 9000.0		& & 5397 &  \cite{NGC925}	\\
NGC 1365	& 66560.0 	& 16060.0 	& & 5972 & \cite{NGC1365}	\\
NGC 1425	& 58800.0 	& 29700.0 	& & 5972/6431  & \cite{NGC1425} \\
NGC 1637 	& 26400.0		& 13200.0      	& & 9155 & \cite{NGC1637-1,NGC1637-2} \\

NGC 2541	& 28760.0		& 12760.0		& & 5972 & \cite{NGC2541} \\
NGC 2841 	& 26400.0		& 11000.0		& & 8322 & \cite{NGC2841} \\

NGC 3031 (M81) & 2000.0		& 2000.0		& & 9073 & \cite{M81}\\

NGC 3198	& 27760.0		& 12560.0		& & 5972 & \cite{NGC3198}	\\
NGC 3319	& 26400.0		& 10400.0		& & 6431 & \cite{NGC3319}	\\
NGC 3351 (M95) 	& 31900.0		& 9830.0 		& & 5397 & \cite{NGC3351}	\\

NGC 3621-1	& 5200.0		& 7800.0		& & 8584 & \cite{NGC3351-OFF}	\\
NGC 3621-2	& 20759.0		& 7380.0		& & 5397 & \cite{NGC3621}	\\
NGC 3627 (M66)   	& 58800.0		& 25000.0		& & 6549 & \cite{NGC3627}	\\
NGC 4321 (M100) 	& 32750.0		& 17150.0		& & 5397 & \cite{NGC4321}	\\
NGC 4414-1	& 1600.0		& 1600.0		& & 8400 & Hubble Heritage	\\
NGC 4414-2	& 32430.0		& 10230.0		& & 5397 & \cite{NGC4414}	\\
NGC 4496A	& 68000.0		& 16000.0		& & 5427 & \cite{NGC4496A}	\\
NGC 4527	& 60000.0		& 25000.0		& & 7504 & \cite{NGC4527}	\\

NGC 4535	& 48800.0 	& 31200.0 	& & 5397/6431 & \cite{NGC4535}	\\
NGC 4536	& 68000.0		& 20000.0		& & 5427 & \cite{NGC4536}	\\
NGC 4548 (M98)	& 48500.0		& 30900.0		& & 6431 & \cite{NGC4548}	\\
NGC 4559		& 2000.0		& 2000.0		& & 9073 & \cite{NGC4559}	\\
NGC 4571	& 10400.0		& 26400.0		& & 6833 & \cite{NGC4571a,NGC4571b}	\\

NGC 4603	& 58800.0		& 14800.0		& & 6439 & \cite{NGC4603}	\\
NGC 4639	& 58800.0		& 13000.0		& & 5981 & \cite{NGC4639}	\\
NGC 4725	& 32430.0 	& 10230.0		& & 5397 & \cite{NGC4725}	\\
NGC 6946	& 2000.0		& 2000.0		& & 9073 & \cite{M51}	\\
NGC 7331	& 40660.0		& 9860.0		& & 5397 & \cite{NGC7331}	\\

UGC 2302	& 15000.0 	& 15600.0		& & 8255 & \cite{UGC2302}	\\
UGC 6614  	& 10100.0 	& 10100.0 	& & 8213 & \cite{UGC6614}	\\
NGC 5194-1 (M51)		& 2000.0		& 2000.0		& & 9073 & \cite{M51}	\\
NGC 5194-2 (M51)		& 2000.0		& 2000.0		& & 9073 & \cite{M51}	\\
	
\enddata

\end{deluxetable}

%
\begin{deluxetable}{l l l l l l l l l l}
\rotate
\tablewidth{0pt.}
\tablecaption{Basic Data}
\tablehead{
\colhead{Galaxy} 	& \colhead{Type}	& \colhead{RA} 	& \colhead{Dec} 	& \colhead{PA}	 	& \colhead{Incl.} 	& \colhead{$D_{25}$} 	& \colhead{Distance.}	& \colhead{Galactic} & \colhead{Remarks} \\ 
\colhead{}			& \colhead{}		& \colhead{}		& \colhead{}		& \colhead{}		& \colhead{}		& \colhead{}			& \colhead{(A)} 		& \colhead{Extinction}	& \colhead{}\\
\colhead{}			& \colhead{}		& \colhead{(deg)}	& \colhead{(deg)}	& \colhead{(deg)}	& \colhead{(deg)}	& \colhead{(arcmin)}	& \colhead{(Mpc)} 		& \colhead{(mag)}	& \colhead{}}
\startdata
NGC 925 		& SAbd 	  	& 36.820469	& 33.578880	& -70	& 62.48	& 11.22 	& 9.16		& 0.147	& 	 \\
NGC 1365	& SBb     		& 53.401909	& -36.140659	& 49		& 34.41	& 10.47	& 17.95		& 0.039	&	 \\
NGC 1425	& SBb	  	& 55.548061	& -29.893511	& -51	& 57.32 	& 5.75	& 21.88		& 0.025	&	 \\
NGC 1637 	& SAB(rs)c 	& 70.367622	& -2.858040	& 27		& 30.68	& 3.98 	& 8.2 		& 0.078	& (1)\\
NGC 2541	& SAcd    		& 123.666969	& 49.061440	& -7		& 56.63 	& 6.31	& 11.22  		& 0.097	&	\\
NGC 2841 	& SAb     		& 140.511063	& 50.976479	& -30	& 60.33	& 8.13	& 14.1 		& 0.030	& (2)\\
NGC 3031 (M81)	& SA(s)ab		& 148.88826	& 69.06526	& -38.5	& 58.87 	&		& 3.63 		& 0.155	&	\\
NGC 3198	& SBc     		& 154.979126	& 45.549690	& 37		& 61.77 	& 8.51	& 13.80		& 0.024	&	\\
NGC 3319	& SB(rs)cd 	&  159.789719	& 41.686871	& 39		& 69.51	& 6.17	& 13.30  		& 0.028	&	\\
NGC 3351 (M95) 	& SBb		& 160.990555	& 11.703610	& -2 		& 14.53	& 7.41	& 10.00 		& 0.054	&	\\
NGC 3621	& SAc 	  	& 169.567917	& -32.812599	& -16	& 58.13	& 12.3	& 6.64    		& 0.156	&	\\
NGC 3627 (M66)   	& SAB(s)b		& 170.062607	& 12.991290	& 1		& 57.38	& 9.12	& 10.05    		& 0.063	&	\\
NGC 4321 (M100) 	& SABbc   	& 185.728745	& 15.822380	& -57	& 39.65	& 7.41	& 15.21    		& 0.051	&	\\
NGC 4414	& SAc     		& 186.612869	& 31.223545	& -20	& 48.70	& 3.63	& 17.70    		& 0.038	&	\\
NGC 4496A	& SBm	  	&  187.913361	& 3.939467	& 65    	& 58.67	& 3.98	& 14.86		& 0.048	&	\\
NGC 4527	& SAB(s)bc	& 188.535400	& 2.653810	& 67    	& 70.73	& 6.17	& 14.1 		& 0.043	& (3)	\\
NGC 4535	& SABc     		& 188.584625	& 8.197760	& 27    	& 28.36	& 7.08	& 15.78   		& 0.038	&	 \\
NGC 4536	& SAB(rs)bc 	& 188.613037	& 2.187880	& -70   	& 62.61	& 7.08	& 14.93   		& 0.035	&     \\
NGC 4548 (M98)	& SBb		& 188.860123	& 14.496320	& -60   	& 22.48	& 5.37	& 16.22   		& 0.074	&     \\
NGC 4559	& SAB(rs)cd	& 188.990372	& 27.959761	& -37	& 61.77	& 10.72	& 10.88		& 0.034	& 	\\
NGC 4571	& SA(r)d  		& 189.234879	& 14.217357	& 75		& 23.07	& 3.63	& 14.9		& 0.091	& (4) \\
NGC 4603	& SA(rs)bc 	& 190.229980	& -40.976402	& 30		& 50.21 	& 3.39	& 33.3 		& 0.325	& (5)	\\
NGC 4639	& SABbc	  	& 190.718140	& 13.257536	& -25	& 43.95 	& 2.75	& 21.98   		& 0.050	&	\\
NGC 4725	& SABab	  	& 192.610886	& 25.500759	& 47		& 45.25	& 10.72	& 12.36   		& 0.023	&	\\
NGC 5194 (M51)	& SA(s)bc 	& 202.46957  	& 47.19526	&  45.0	& 25.58 	& 11.22	& 8.4 		& 0.067	&	\\
NGC 6946 	& SAB(rs)cd	& 308.718048	& 60.153679	& 70		& 39.65	& 11.48	& 11.48		& 0.663	&	\\
NGC 7331	& SAb	  	& 339.267090	& 34.415920	& -12	& 57.38	& 10.47	& 14.72   		& 0.176	&	\\
UGC 2302	& SB(rs)m		& 42.285831	& 2.127265	& 60.0	& 1.0		& 4.79	& 14.7 		& 0.156	& (6)	\\
UGC 6614  	& SA(r)a 		&174.811844	& 17.143578	& -65 	& 40.54 	& 1.66 	& 84.68		& 0.055	& (7)	\\
\enddata
\tablenotetext{A}{All distances were taken from \cite{KeyProject} with some exceptions, noted below.}
\tablenotetext{1}{\cite{NGC1637-1} and \cite{NGC1637-2}.} 
\tablenotetext{2}{\cite{NGC2841}.}
\tablenotetext{3}{\cite{NGC4527}.}
\tablenotetext{4}{\cite{NGC4571a} and \cite{NGC4571b}.}
\tablenotetext{5}{\cite{NGC4603}}
\tablenotetext{6}{No distance from Cepheid method or supernovae available. Using NED data (1104 ${\rm km \ s^{-1}}$) and ${\rm H_0 = 75 \ km^{-1} \ Mpc^{-1}}$}
\tablenotetext{7}{No distance from Cepheid method or supernovae available. Using NED data (6351 ${\rm km \ s^{-1}}$) and ${\rm H_0 = 75 \ km^{-1} \ Mpc^{-1}}$}
\end{deluxetable}

\begin{deluxetable}{l l l l l l l l l l l l}
\rotate
\tablewidth{0pt.}
\tablecaption{Radial Extinction per field (mag)}
\tablehead{
		&	&	&	& \colhead{${\rm R/R_{25}}$} & & & &   \\	 
\colhead{Galaxy} 	& \colhead{0-0.25} & \colhead{0.25-0.5} & \colhead{0.5-0.75} & \colhead{0.75-1.0} & \colhead{1.0-1.25} & \colhead{1.25-1.5} & \colhead{1.5-1.75} & \colhead{1.75-2.0} & \colhead{2.0-2.25} & \colhead{2.25-2.5} & \colhead{2.5-2.75} \\  }	
\startdata

NGC 925  		& $-0.1^{+1.7}_{-1.5}$	& $-0.7^{+0.6}_{-0.6}$		& $-0.4^{+0.5}_{-0.5}$		& $1.1^{+1.1}_{-1.4}$  	 	& \nodata   		& \nodata   		& \nodata   	& \nodata   			& \nodata & \nodata & \nodata \\ 
NGC 1365     	& \nodata   		& $ 0.8^{+0.6}_{-0.6}	$ 	& $ 0.3^{+0.4}_{-0.4} $ 	& \nodata				& \nodata   		& \nodata   		& \nodata   		& \nodata   		& \nodata & \nodata & \nodata \\ 
NGC 1425     	& \nodata   		& \nodata				& $0.8^{+1.1}_{-1.2}$   	& $ 1.1^{+0.7}_{-0.7}$   	& $0.1^{+0.4}_{-0.4}$	& $0.6^{+0.6}_{-0.6}$	& $0.6^{+1.1}_{-1.4}$   & \nodata   		& \nodata & \nodata & \nodata \\ 
NGC 1637     	& \nodata			& $ 2.9^{+1.6}_{-2.0} $   	& $2.2^{+1.2}_{-1.6} $  	& $ 1.1^{+1.1}_{-1.2} $   	& $0.5^{+1.1}_{-1.2}$	& \nodata 			& \nodata   		& \nodata   		& \nodata & \nodata & \nodata \\ 
NGC 2541     	& $2.2^{+1.7}_{-2.4}$	& $ 0.6^{+0.8}_{-0.9} $   	& $1.0^{+0.6}_{-0.6} $  	& $ 0.4^{+0.4}_{-0.5} $   	& \nodata  & \nodata   		& \nodata   		& \nodata   		& \nodata & \nodata & \nodata \\ 
NGC 2841     	& $-0.3^{+3.5}_{-3.0}$	& $ 1.3^{+1.0}_{-1.2} $   	& $1.9^{+1.0}_{-1.2} $  	& $ -0.3^{+0.9}_{-0.9} $  	& $-0.5^{+1.8}_{-1.5}$	& \nodata  	 	& \nodata   		& \nodata   		& \nodata & \nodata & \nodata \\ 
NGC 3031	& $1.7^{+0.9}_{-1.1}$ 	& $ 0.7^{+0.9}_{-1.0}$ 		& \nodata	 			& \nodata   			& \nodata   		& \nodata   		& \nodata   		& \nodata   		& \nodata & \nodata & \nodata \\ 
NGC 3198     	& \nodata			& $ 1.6^{+0.9}_{-1.0} $  	& $0.4^{+0.5}_{-0.5} $  	& $ 0.9^{+0.7}_{-0.7} $   	& \nodata   		& \nodata   		& \nodata   		& \nodata   		& \nodata & \nodata & \nodata \\ 
NGC 3319     	& $1.5^{+1.2}_{-1.4}$	& $ 1.7^{+0.9}_{-1.1} $   	& $0.5^{+0.5}_{-0.5} $  	& $ -0.3^{+1.6}_{-1.5} $  	& \nodata   		& \nodata   		& \nodata   		& \nodata   		& \nodata & \nodata & \nodata \\ 
NGC 3351  	& 				& $ 0.6^{+0.9}_{-1.1} $   	& $0.7^{+1.2}_{-1.9} $  	& \nodata  	& $0.8^{+1.5}_{-1.7}$	& $-0.1^{+0.8}_{-0.8}$	& $3.0^{+2.0}_{-3.8}$	& $0.9^{+1.4}_{-1.7} $	& \nodata	& $0.0^{+3.0}_{-3.0}$   & \nodata   \\
NGC 3621	    	&  $-0.3^{+3.7}_{-3.0}$ &$1.6^{+0.5}_{-0.5} $		& $0.9^{+0.6}_{-0.7}$		& \nodata   			& \nodata   		& \nodata   		& \nodata   		& \nodata   		& \nodata & \nodata & \nodata \\  
NGC 3621-1     	& \nodata   		& $2.2^{+0.7}_{-0.7}$		& \nodata 				& \nodata   			& \nodata   		& \nodata 			& \nodata   	& \nodata   			& \nodata & \nodata & \nodata \\ 
NGC 3621-2	& $-0.3^{+3.7}_{-3.0} $ & $1.2^{+0.5}_{-0.5} $  	& $0.7^{+0.6}_{-0.6} $  	& \nodata   			& \nodata  		& \nodata 			& \nodata   	& \nodata   			& \nodata & \nodata & \nodata \\ 
NGC 3627  	& \nodata		 	& $3.2^{+1.4}_{-1.6}$		& $1.1^{+0.7}_{-0.8}$   	& \nodata				& \nodata   		& \nodata   		& \nodata   	& \nodata   			& \nodata & \nodata & \nodata \\ 
NGC 4321 	& \nodata   		& $1.2^{+1.4}_{-1.6}$		& $3.1^{+1.4}_{-1.8}$	   	& $2.7^{+1.4}_{-2.1}$   	& \nodata  		& \nodata   		& \nodata   	& \nodata   			& \nodata & \nodata & \nodata \\ 
NGC 4414 	& \nodata	    		& $ 0.4^{+1.6}_{-1.7}$ 		& $2.1^{+1.0}_{-1.1}$		& $0.3^{+0.4}_{-0.4}$		& $1.1^{+0.6}_{-0.7}$	& $0.6^{+0.8}_{-0.9}$    & \nodata   	& \nodata   			& \nodata & \nodata & \nodata \\ 
NGC 4414-1	& \nodata   		& $0.2^{+2.3}_{-2.4}$		& $4.8^{+2.5}_{-4.9}$  		& $0.1^{+0.6}_{-0.6}$   	& $2.7^{+1.7}_{-3.2}$	& $0.2^{+1.0}_{-1.1}$	& \nodata 		& \nodata   		& \nodata & \nodata & \nodata \\ 
NGC 4414-2	& \nodata   		& $0.6^{+2.2}_{-2.4}$		& $1.2^{+1.1}_{-1.2}$  	 	& $0.5^{+0.6}_{-0.6}$   	& $0.7^{+0.6}_{-0.7}$	& $1.1^{+1.4}_{-1.9}$	& \nodata   	& \nodata   			& \nodata & \nodata & \nodata \\ 
NGC 4496A 	& \nodata		& $5.1^{+2.0}_{-2.2}$   	& $1.0^{+0.7}_{-0.7}$   	& $1.1^{+1.3}_{-1.6}$   	& \nodata  		& \nodata    		& \nodata   	& \nodata   			& \nodata & \nodata & \nodata \\ 
NGC 4527     	& $2.8^{+4.4}_{-7.1} $	& $5.7^{+2.9}_{-5.5}$   	& $2.1^{+1.0}_{-1.0}$   	& $0.4^{+0.4}_{-0.5}$   	& \nodata 			& \nodata   		& \nodata   	& \nodata   			& \nodata & \nodata & \nodata \\ 
NGC 4535     	& $-0.6^{+1.1}_{-0.8} $	& $1.4^{+1.1}_{-1.3}$  	 	& $1.2^{+0.9}_{-1.0}$   	& $0.6^{+0.5}_{-0.5}$   	& $-0.2^{+1.6}_{-1.6}$	& \nodata   		& \nodata   	& \nodata   			& \nodata & \nodata & \nodata \\ 
NGC 4536     	& \nodata   		& $0.0^{+0.6}_{-0.5}$		& $1.0^{+0.6}_{-0.6}$ 		& $0.9^{+0.6}_{-0.7}$   	& \nodata 			& \nodata   		& \nodata   	& \nodata   			& \nodata & \nodata & \nodata \\ 
NGC 4548 (M98)   	& \nodata   		& $-4.2^{+26.1}_{-13.6}$	& $-0.1^{+0.7}_{-0.7}$		& $2.6^{+1.5}_{-2.3}$   	& $1.1^{+1.1}_{-1.2}$	& $0.9^{+0.9}_{-1.0}$	& $0.9^{+1.4}_{-1.5}$   &$ 0.9^{+1.0}_{-1.1} $ &$ 0.6^{+0.7}_{-0.8} $   &$ -1.0^{ +1.6}_{-1.2}$   &  \nodata   \\
NGC 4559     	& $0.4^{+1.0}_{-1.1}$   & $0.0^{+0.6}_{-0.6}$		& $0.3^{+0.4}_{-0.5}$		& $-1.0^{+1.4}_{-1.0}$  	& \nodata   		& \nodata   		& \nodata   	& \nodata   			& \nodata & \nodata & \nodata \\ 
NGC 4571     	& $2.8^{+2.0}_{-2.7}$   & $1.4^{+0.7}_{-0.8}$		& $0.9^{+0.7}_{-0.7}$		& $1.1^{+0.8}_{-1.0}$   	& \nodata 			& \nodata   		& \nodata   	& \nodata   			& \nodata & \nodata & \nodata \\ 
NGC 4603     	& $3.6^{+1.1}_{-1.3}$   & $0.8^{+0.5}_{-0.5}$		& $0.3^{+0.7}_{-0.8}$		& \nodata   			& \nodata   		& \nodata   		& \nodata   	& \nodata   			& \nodata & \nodata & \nodata \\ 
NGC 4639     	& $9.4^{+3.6}_{-5.7}$   & $0.3^{+0.5}_{-0.5}$		& $0.5^{+0.5}_{-0.5}$		& $-0.2^{+1.0}_{-1.0}$		& \nodata   		& \nodata   		& \nodata   	& \nodata   			& \nodata & \nodata & \nodata \\ 
NGC 4725     	& \nodata   		& $1.5^{+1.7}_{-2.9}$		& $0.8^{+0.5}_{-0.6}$		& $0.7^{+0.5}_{-0.5}$		& \nodata		 	& \nodata   		& \nodata   	& \nodata   			& \nodata & \nodata & \nodata \\ 
NGC 5194 	& \nodata			& $1.4^{+1.0}_{-1.1} $		& $1.3^{+0.6}_{-0.6}$		& $0.8^{+1.0}_{-1.2} $  	& \nodata  		& \nodata  		& \nodata  		& \nodata  		& \nodata & \nodata & \nodata \\    
NGC 5194-1	& \nodata  		& $1.3^{+1.1}_{-1.3} $ 		& $ 1.0^{+1.0}_{-1.1} $ 	& $1.0^{+1.1}_{-1.4} $ 		& \nodata 			& \nodata  		& \nodata   		& \nodata   		& \nodata & \nodata & \nodata \\ 
NGC 5194-2	& \nodata 			& $1.4^{+1.6}_{-1.8} $		& $ 1.5^{+0.7}_{-0.8} $  	& $0.8^{+13.2}_{-14.5} $	& \nodata 			& \nodata   		& \nodata   		& \nodata   		& \nodata & \nodata & \nodata \\ 
NGC 6946     	& \nodata   		& $1.4^{+1.5}_{-2.0}$		& $1.8^{+1.2}_{-1.6}$		& $-0.1^{+1.0}_{-1.0}$		& $2.1^{+1.3}_{-2.0}$   & \nodata   		& \nodata   	& \nodata   			& \nodata & \nodata & \nodata \\ 
NGC 7331     	& \nodata   		& $0.3^{+0.8}_{-0.8}$		& $0.2^{+0.4}_{-0.4}$		& $1.0^{+0.7}_{-0.8}$		& \nodata   		& \nodata   		& \nodata   	& \nodata   			& \nodata & \nodata & \nodata \\ 

UGC 2302     	& $0.5^{+1.0}_{-1.1}$	& $0.5^{+0.6}_{-0.6}$		& $0.6^{+0.4}_{-0.4}$		& $1.2^{+1.3}_{-1.6}$   	& \nodata   		& \nodata   		& \nodata   	& \nodata   			& \nodata & \nodata & \nodata \\ 
UGC 6614     	& \nodata   		& \nodata   			& $0.4^{+2.0}_{-2.5}$		& $0.6^{+0.9}_{-1.2}$   	& $0.6^{+1.1}_{-1.2}$	& $0.5^{+1.1}_{-1.2}$	& $-0.6^{+0.8}_{-0.8}$   & $-0.3^{+0.7}_{-0.7}$& $0.2^{+0.7}_{-0.7}$ & $0.6^{+0.6}_{-0.7}$   & $0.5^{+0.7}_{-0.7}$ \\ 
\enddata
\tablecomments{The individual radial extinction measurements of the galaxies in our sample based on the field galaxy counts. 
Note that the intrinsic uncertainty in the background field of galaxies keeps the error in these measurements high. The counts 
have been corrected for the difference in Galactic Extinction between the target foreground galaxy and the average of the HDF 
fields. These values are {\it not} corrected for incliination in any way. UGC6614 has the largest radial coverage due to its extreme distance; in addition to the values shown, UGC6614 has an 
extinction measurement for the radial interval between 2.75 and 3 $R_{25}$ of $0.8^{+1.2}_{-1.3}$.
The averages presented in the rest of the paper are the result of combining the numbers of background galaxies and we caution 
against averaging these values to get a profile for a subset of the sample.}
\end{deluxetable}

\clearpage


\begin{figure}
\plotone{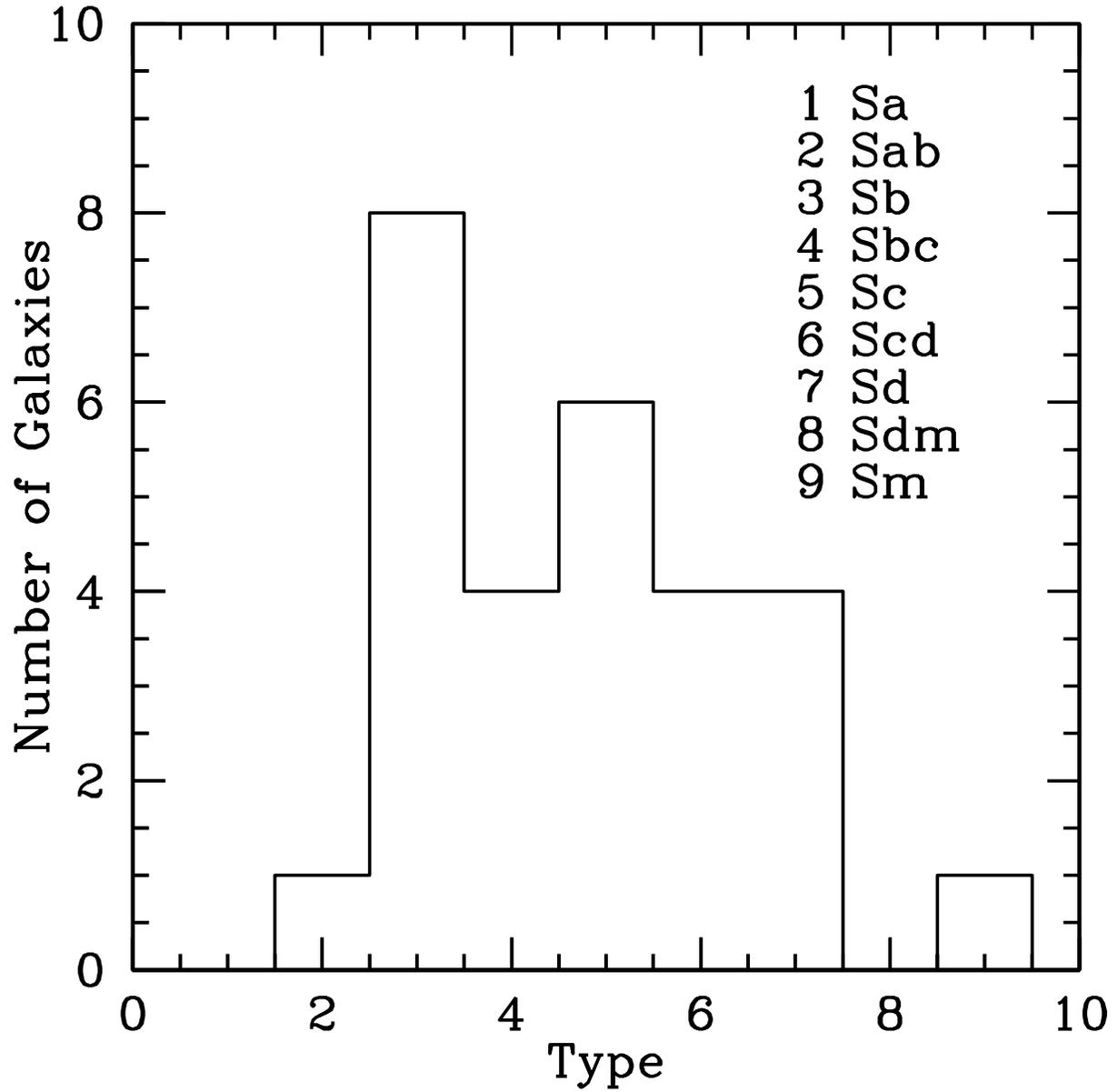}
\caption{The distribution of Hubble type \citep{RC3} for our HST sample. 
The Distance Scale Key Project \citep{KeyProject}, from which most of our sample is drawn, 
concentrated on later types to maximise the number of Cepheids.}
\end{figure}


\begin{figure}
\plotone{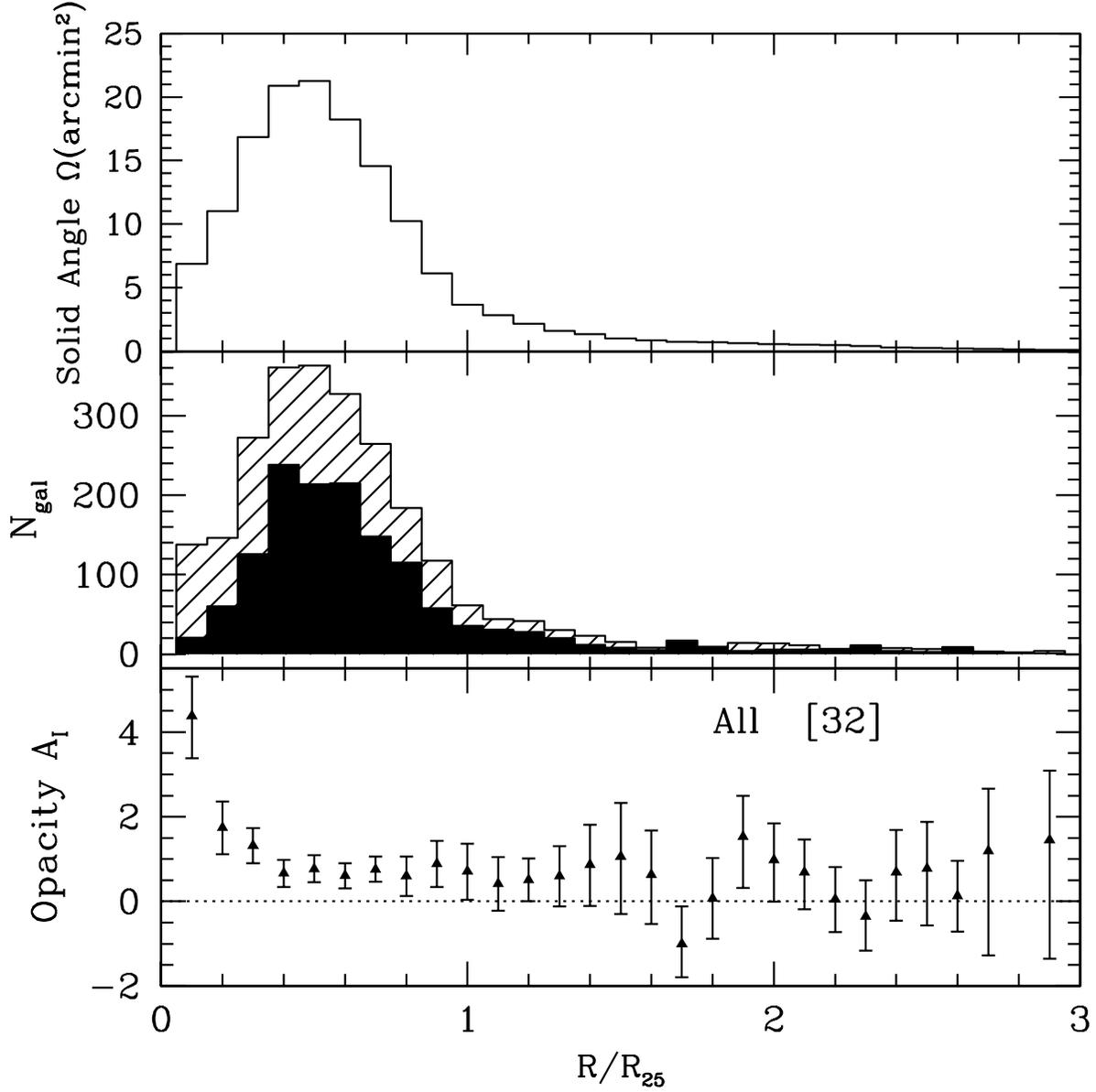}
\caption{The composite of our entire sample (32 WFPC2 fields). The top panel shows the total solid angle in each 
annulus  as a function of scaled radius. The number of field galaxies found (middle panel) are 
presented for both the synthetic fields without dimming (shaded) and the science fields (solid 
histogram). The bottom panel shows the derived opacity in each annulus as a function of radius. 
No inclination correction has been applied to these results.}
\end{figure}


\begin{figure}
\rotate
\plotone{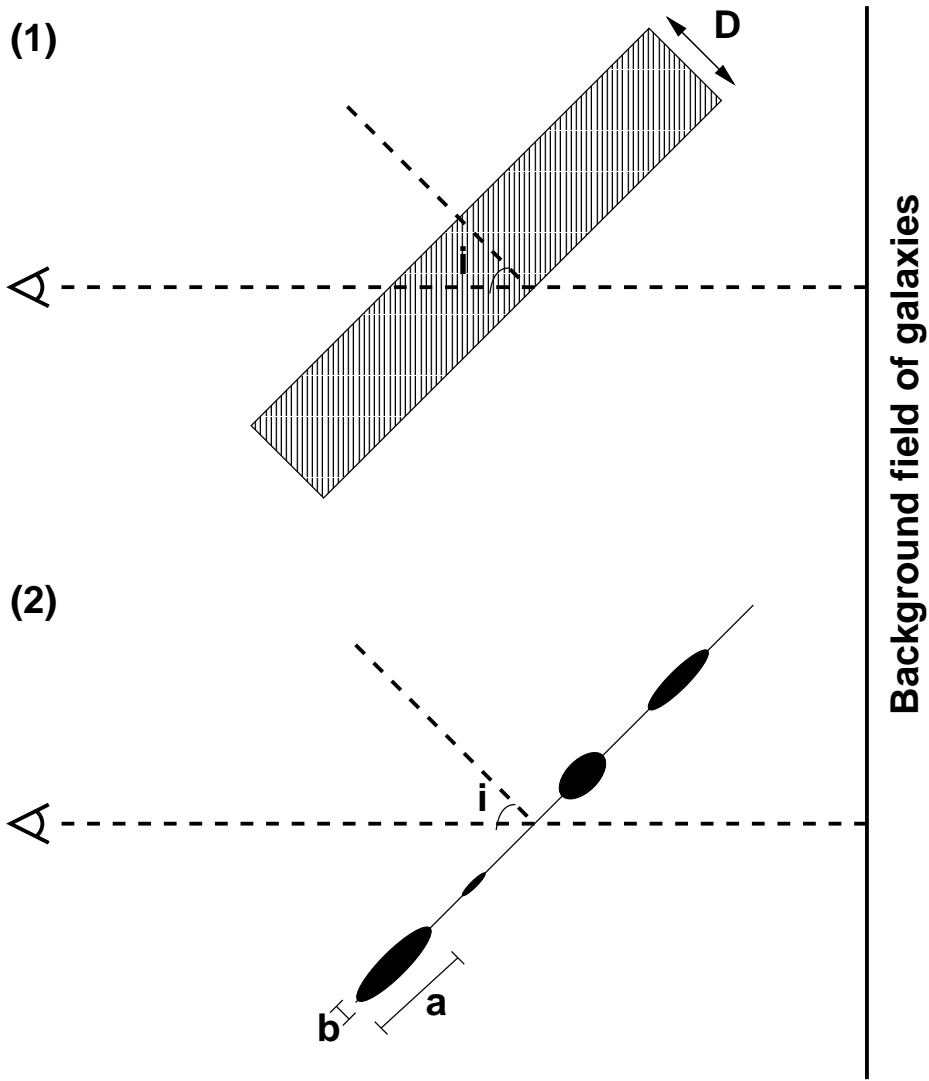}
\caption{Two models of dust geometry in the disk. The flat screen with a thickness D (1) and the screen of 
dark clouds (2). See the text for the effect of the thick screen. The effect of oblong clouds depends on the observed filling factor of the clouds. The observed optical depth ($\tau$) is related to the filling factor by: $\tau = ln(1-f)$. the observed filling factor is related to the face-on value as follows in case of the oblong clouds: $f = \epsilon ~ f_{obs} + (1-\epsilon) ~ cos(i)  ~ f_{obs}$. From the relation between opacity and optical depth ($A=-2.5~log(e^{-\tau}$), equation 1 ($A=-2.5log(N/N_0)$) and these expressions for the filling factor, equation 3 can be derived. The average oblateness ($ \epsilon = 1 - {b \over a}$) of the clouds influences directly the inclination correction. If they are spherical ($\epsilon = 0$), then the effect of inclination on the number of 
field galaxies is most profound. However if they are effectively flat ($\epsilon = 1$), then there is no effect 
of inclination on the numbers (the projection effects on the effective cloud size and filling factor cancel each other).}
\end{figure}


\begin{figure}
\plotone{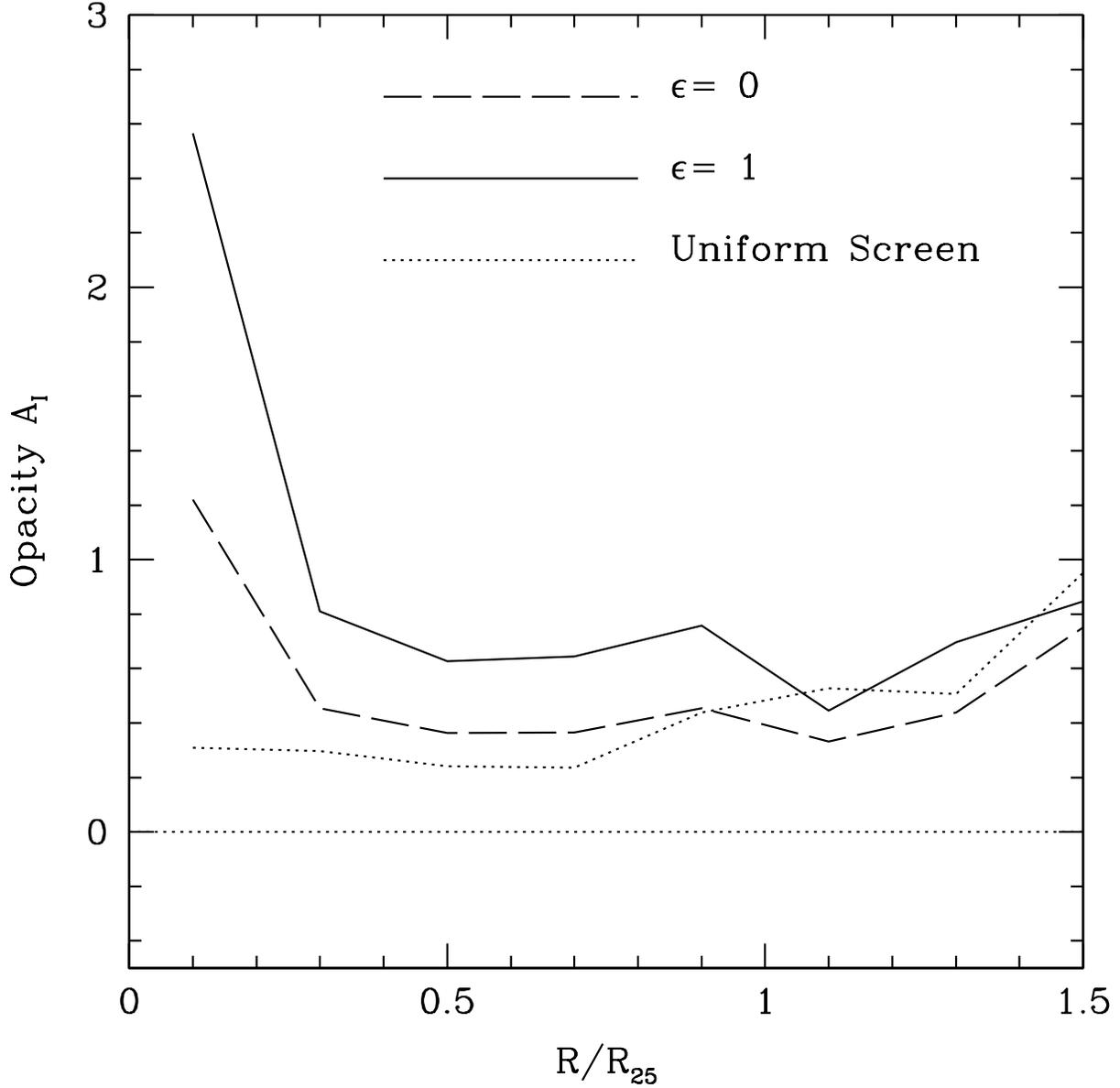}
\caption{The average opacity as a function of radius, derived from the number of field galaxies corrected for inclination with equation 2. Flat clouds ($\epsilon = 1$) do not influence the numbers. The maximum correction ($\epsilon = 0$) has the largest effect on high opacities. The opacity profile from the number of field galaxies corrected for inclination using a smooth screen is also shown.}
\end{figure}
\clearpage

\begin{figure}
\plotone{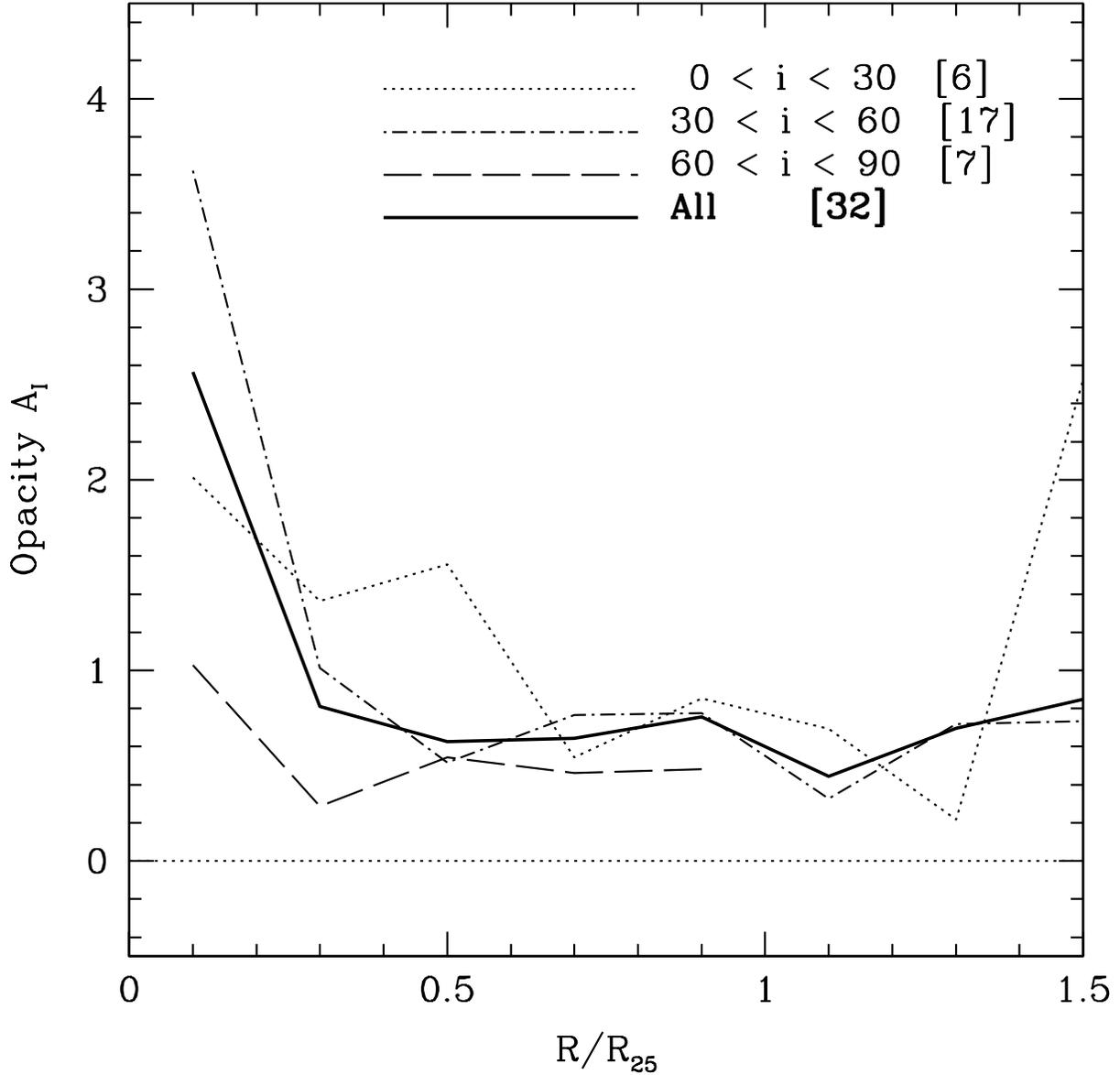}
\caption{The average opacity as a function of radius, taken over our entire sample (thick line) 
and four subsets based on inclination. The number between brackets denotes the number of 
fields in each bin. As there is not discernible trend with inclination, other effects must dominate 
the average opacity. For this reason we choose to ignore the effects of inclination on our measurements. Beyond 1.3 ${\rm R_{25}}$, the values are from poor statistics which explains the occasional negative value.}
\end{figure}
\clearpage


\begin{figure}
\plotone{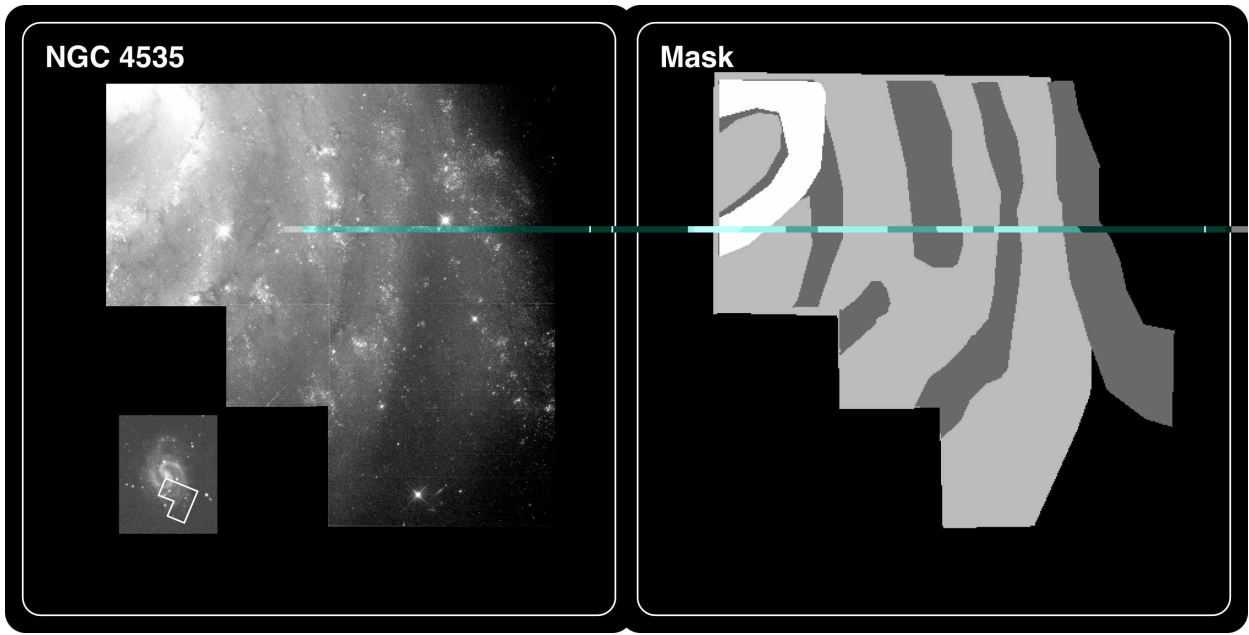}
\caption{The mask of typical regions for NGC4535. White regions represent ``crowded'' regions, 
dark grey are ``arm'' regions, light grey disk regions, ``inter-arm'', and all objects not in any 
of the above categories are disk region ( ``outside'').}
\end{figure}
\clearpage


\begin{figure}
\plotone{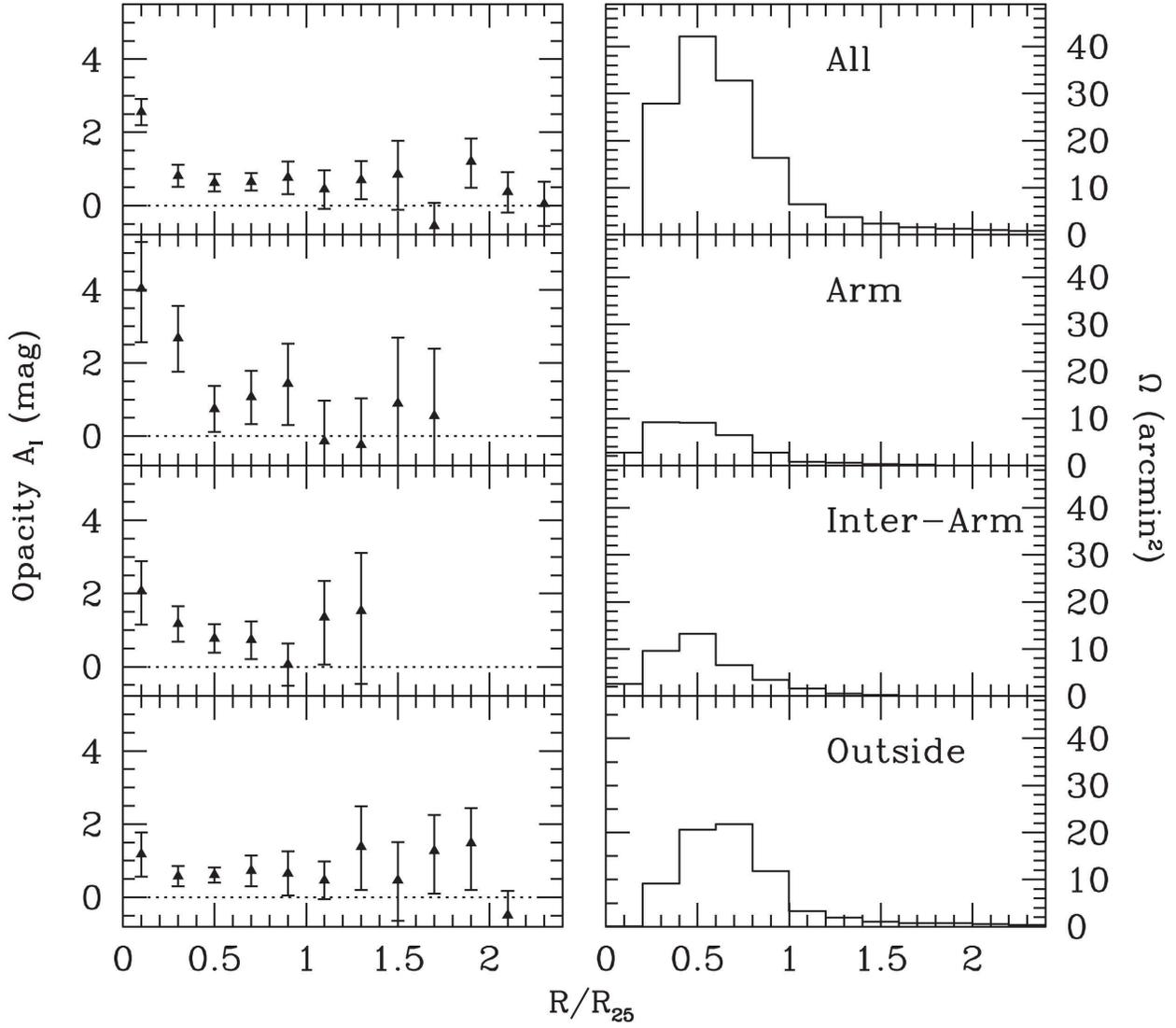}
\caption{The average opacity as a function of radius (left panels), taken over our entire sample for each of three typical regions in the spiral disk: arm regions, disk regions enclosed by spiral arms (inter-arm) and disk regions not enclosed by spiral arms (outside). Right panels show the solid angle as a function of radius.} 
\end{figure}


\begin{figure}
\plotone{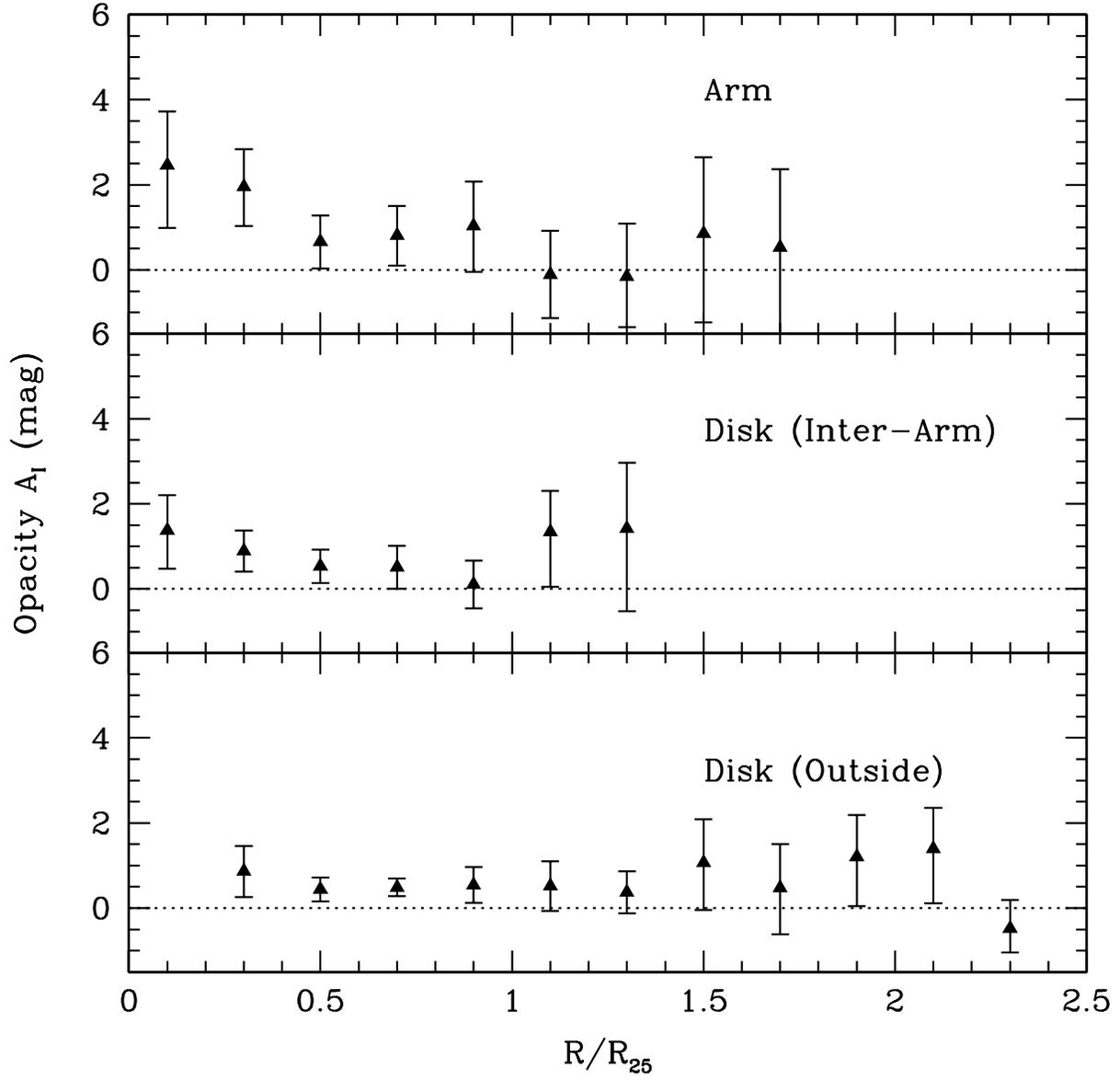}
\caption{Same as Figure 7 but numbers of galaxies from the science fields were corrected for inclination using equation 2 and $\epsilon = 0.5$. }
\end{figure}


\begin{figure}
\plotone{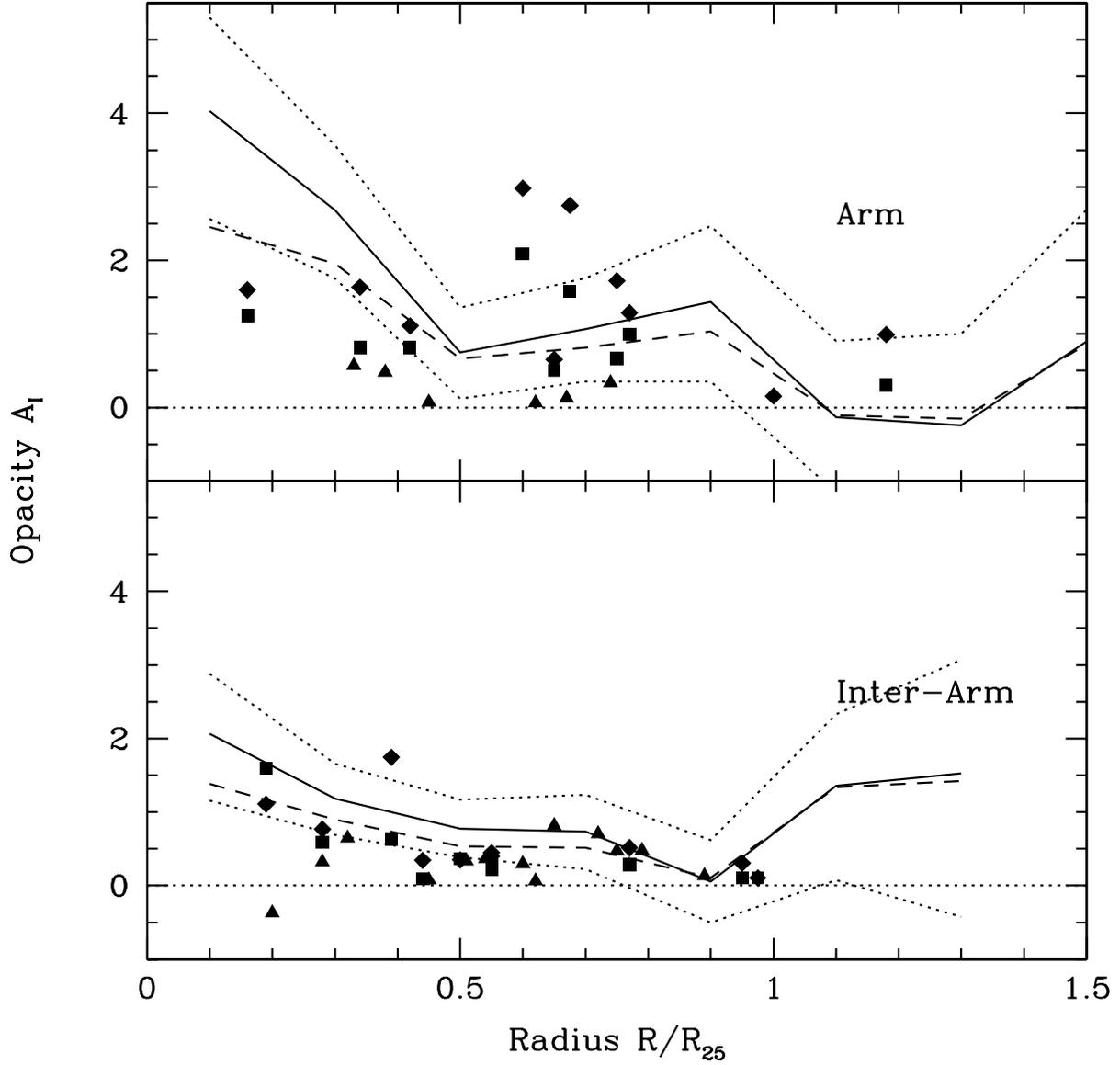}
\caption{The radial extinction profile from the counts of field galaxies (lines) and the occulting 
galaxy technique (points). Top panel shows the ``arm'' regions, bottom the ``inter-arm'' regions. 
The solid line is the SFM extinction profile uncorrected for inclination with the uncertainty 
denoted by the dot-dash lines. The dashed line shows the opacity corrected using $\epsilon = 0.5$. 
The filled squares and diamonds are the $A_I$ and $A_B$ from  \cite{kw00a}, respectively, and the 
triangles are the opacities from \cite{kw00b}. All symbols are uncorrected for inclination. Typical 
uncertainties for these are of the order of a couple of tenths of magnitude. } 
\end{figure}


\begin{figure}
\plotone{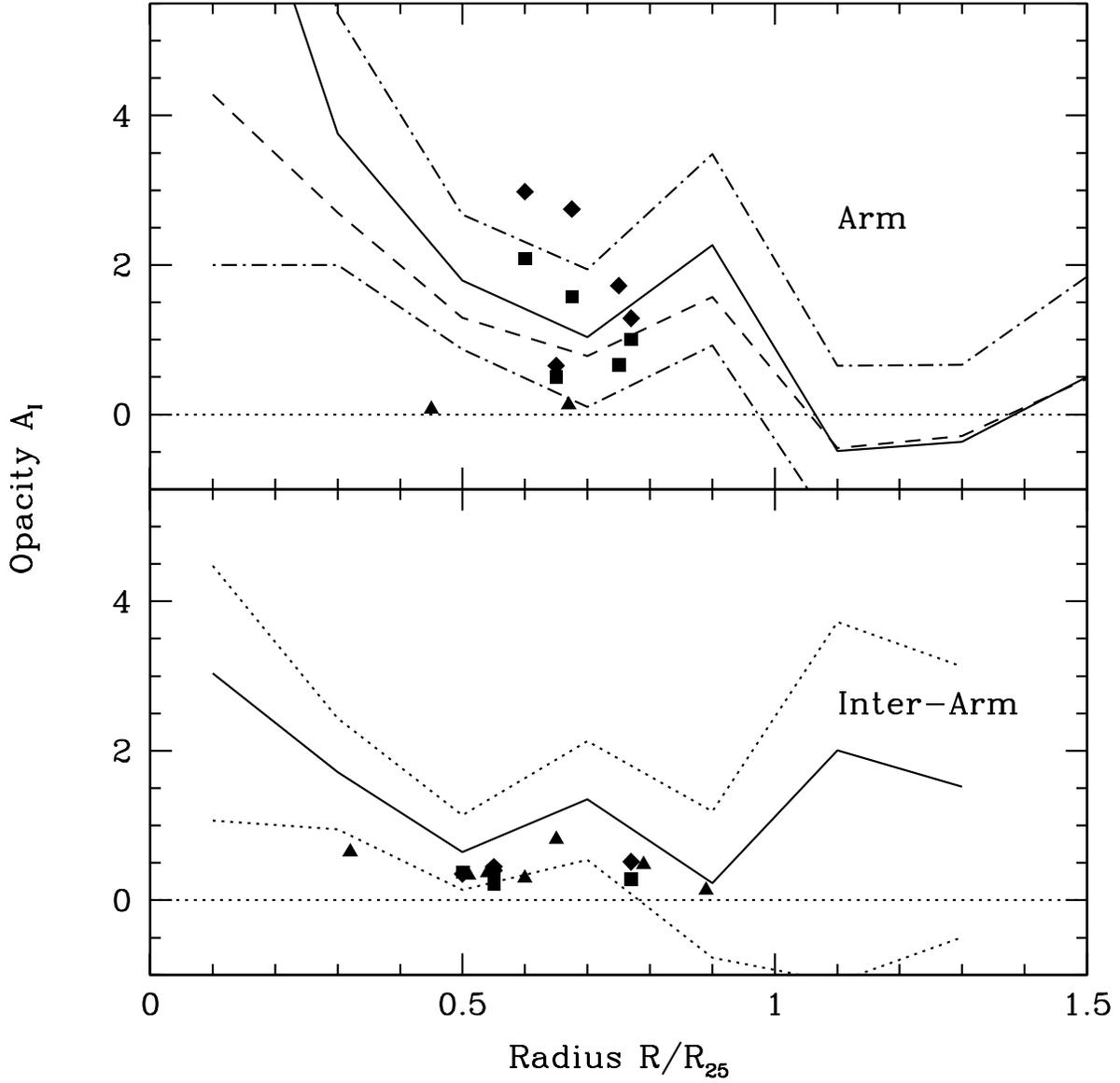}
\caption{The radial extinction profile from the counts of field galaxies and the occulting galaxy 
technique (lines and points as in Figure 9) for the spiral galaxy Hubble subtypes that both 
techniques have in common, the Sb and Sbc galaxies. Points in Figure 9 without a spiral galaxy 
subtype noted in \cite{kw00a} and \cite{kw00b} are omitted. Lines are the same as for Figure 9.}
\end{figure}

\begin{figure}
\plotone{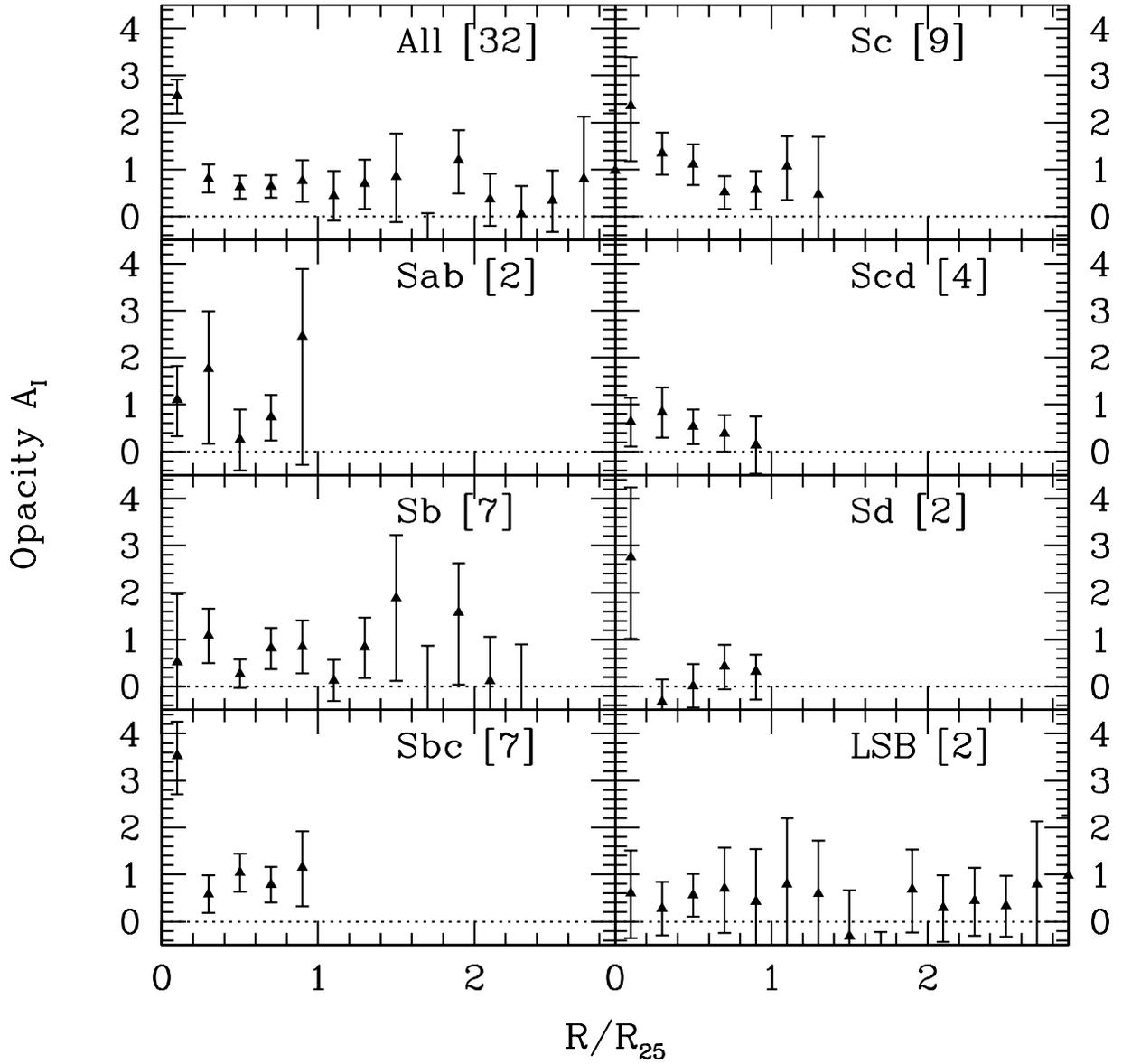}
\caption{The average opacity as a function of radius, taken over our entire sample (top left) and 
for all Hubble types in our sample. LSB galaxies are treated as a separate Hubble type. UGC 6614 
is the most distant galaxy in our sample (with the smallest $R_{25}$), hence the wide coverage 
in radius for the LSB galaxies but with large uncertainties for the opacities. The number between 
brackets is the number of WFPC2 fields averaged for each plot.}
\end{figure}
\clearpage


\begin{figure}
\plotone{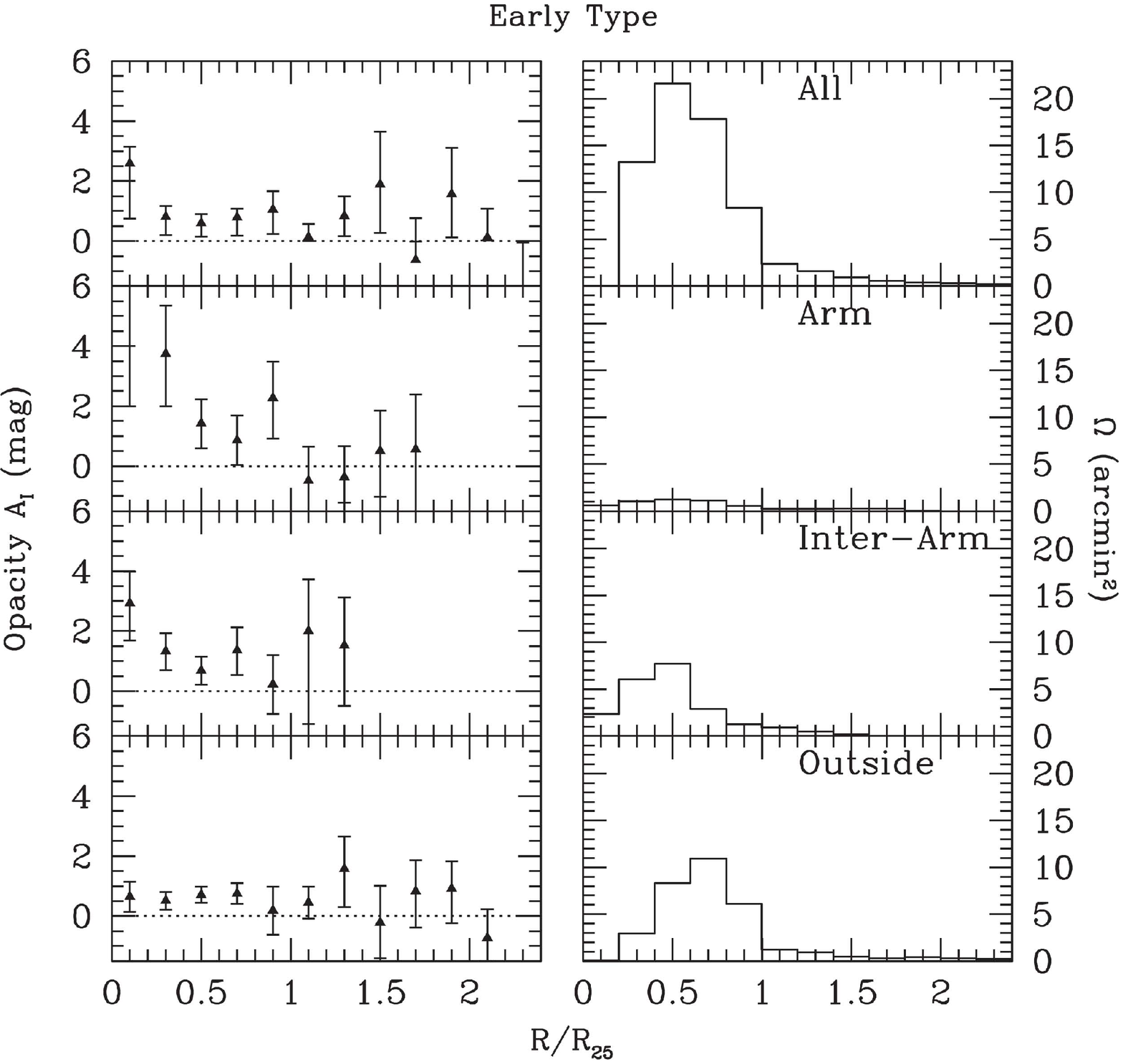}
\caption{The average opacity as a function of radius, taken over the early spirals (Sab,Sb and Sbc) 
in our sample, for each of three typical regions in the spiral disk: arm regions, disk regions enclosed 
by spiral arms (inter-arm) and disk regions not enclosed by spiral arms (outside).} 
\end{figure}


\begin{figure}
\plotone{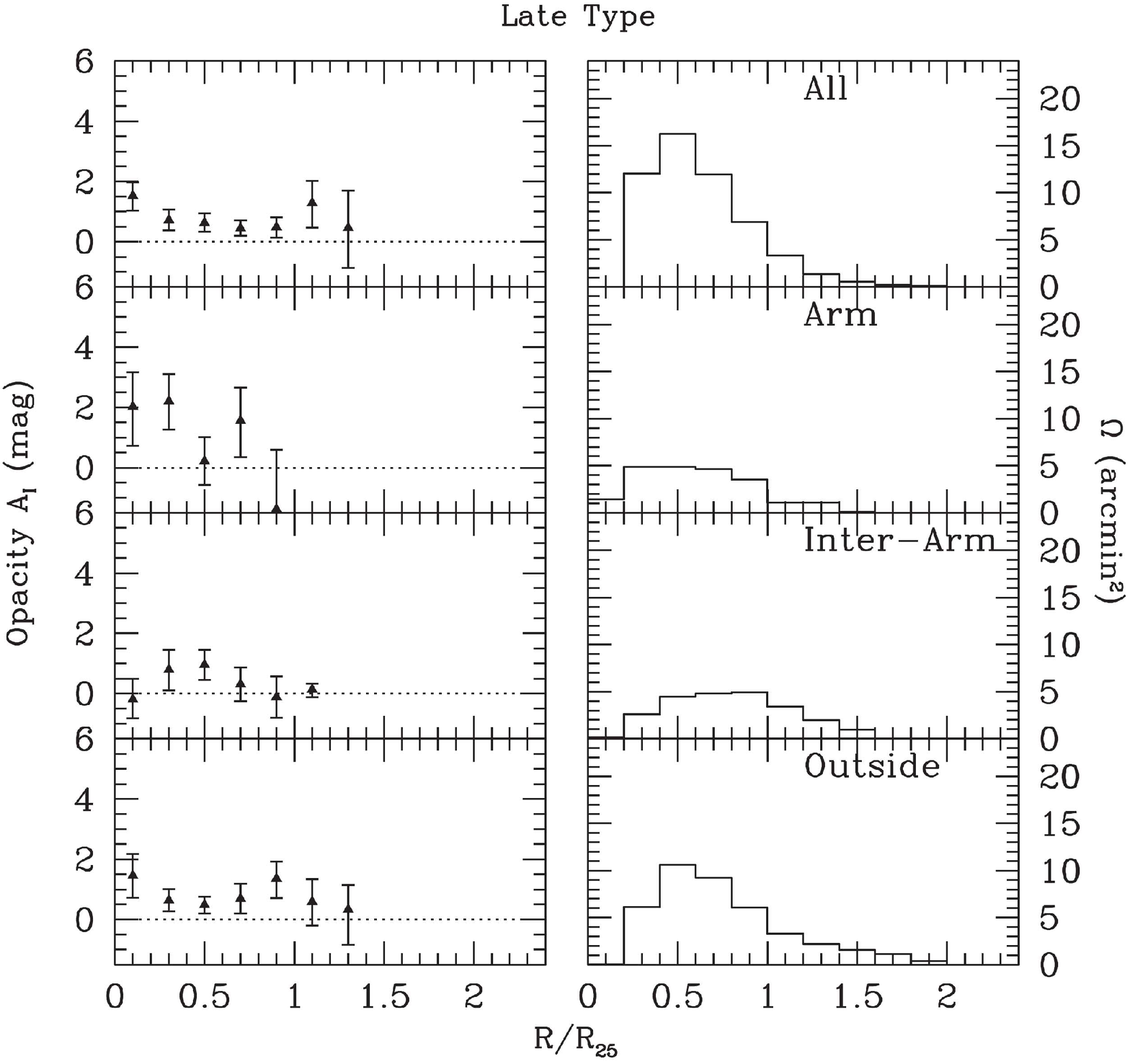}
\caption{The average opacity as a function of radius, taken over the late spirals (Sc,Scd and Sd) in 
our sample, for each of three typical regions in the spiral disk: arm regions, disk regions enclosed by 
spiral arms (inter-arm) and disk regions not enclosed by spiral arms (outside).} 
\end{figure}


\begin{figure}
\plotone{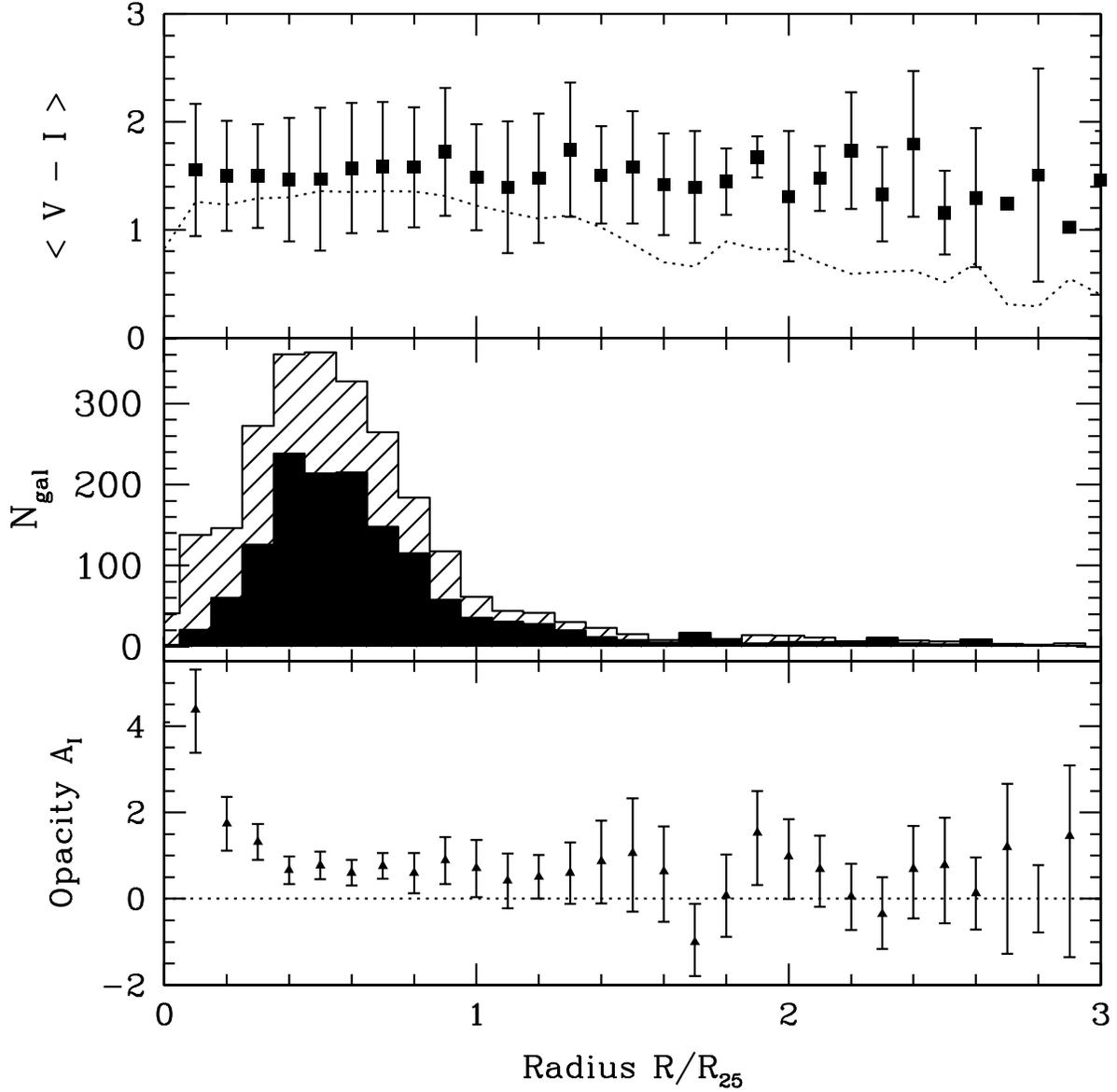}
\caption{Color changes with radius based on the entire sample of 32 WFPC2 exposures. Top panel: 
the average ($V-I$) color of field galaxies found in the science fields as a function of radius. The error 
bars denote the standard deviation of the distribution of ($V-I$) colors. The dotted line is the average of 
the synthetic field objects. Middle panel: the number of field galaxies found in the simulated fields 
without extinction (shaded) and in the science field (filled) as a function of radius. Bottom: the opacity 
derived from the numbers of field galaxies as a function of radius, expressed in $R_{25}$.}
\end{figure}


\begin{figure}
\plotone{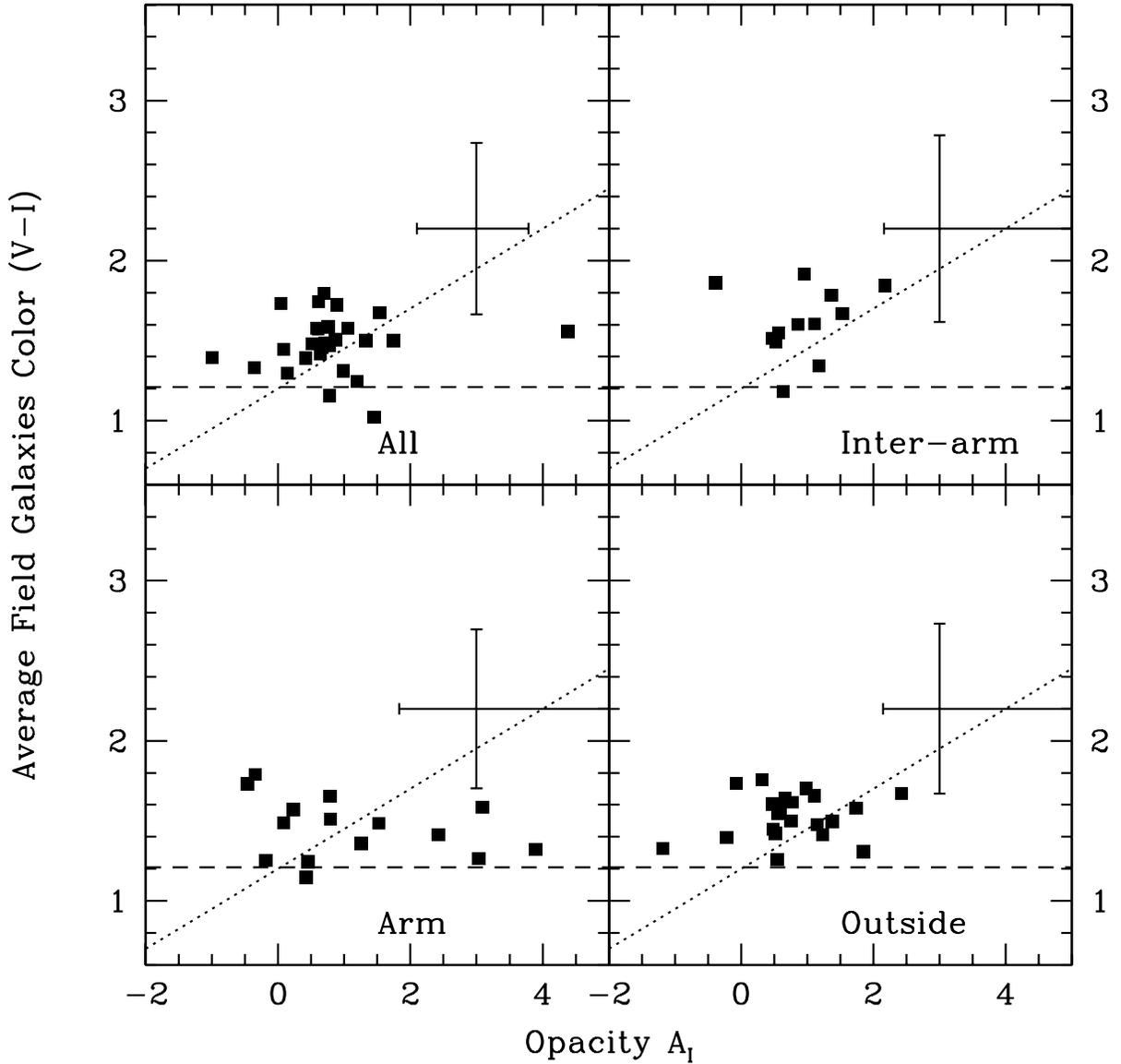}
\caption{The average galaxy color ($V-I$) as a function of average opacity ($A_I$). The error bars denote 
the average uncertainty in opacity and average standard deviation in the distribution of colors of field 
galaxies from the science fields. The dotted line is the Galactic reddening law, normalized on the 
average color of the HDF galaxies. The dashed line the average color of the HDF galaxies, identified 
as such by our algorithm without any foreground field. Selection effects and blends most likely account 
for the reddening compared to the HDF galaxies.  }
\end{figure}


\begin{figure}
\plotone{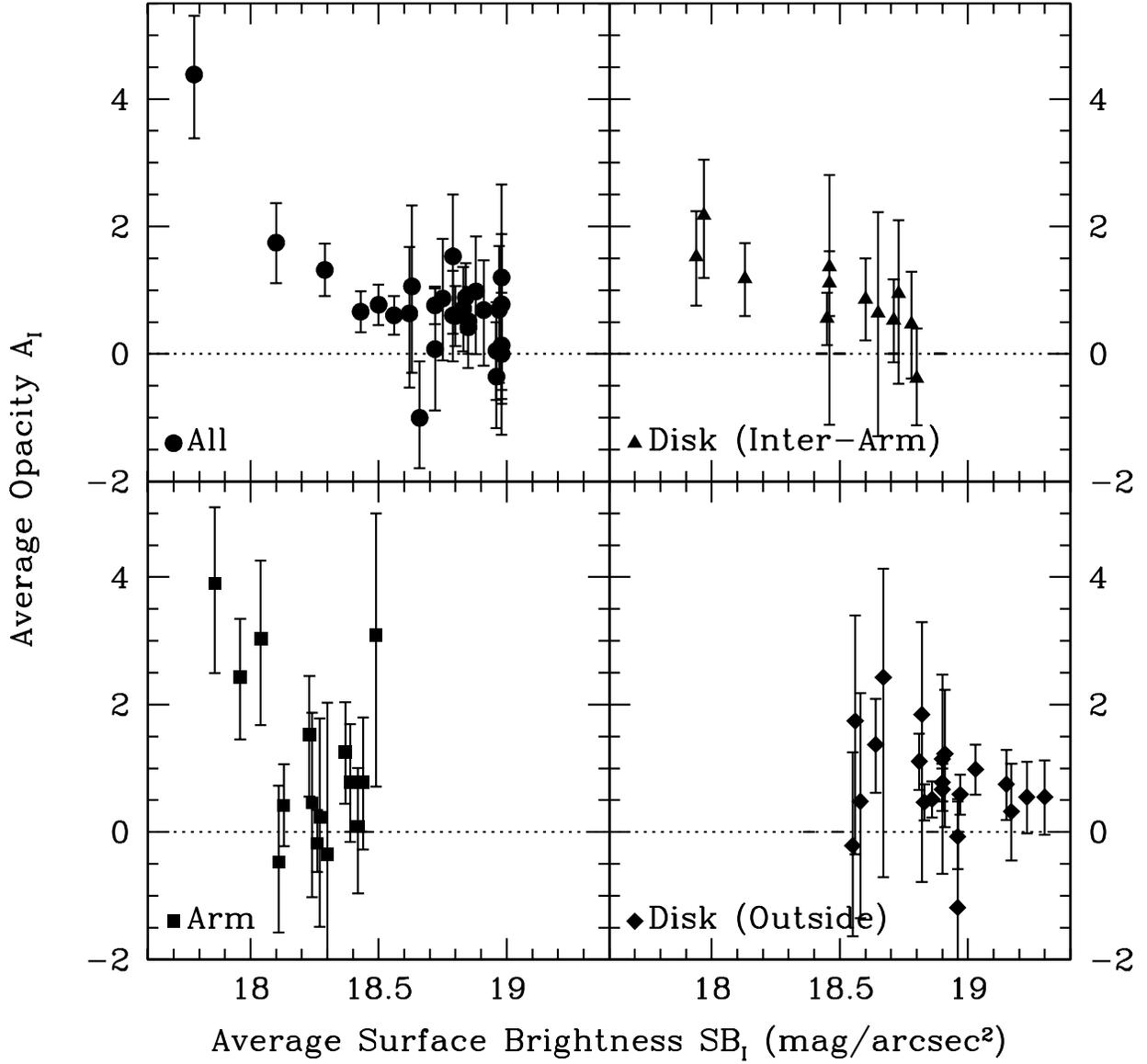}
\caption{The average opacity as a function of average surface brightness. Both surface brightness and 
opacity were determined from the radial annuli in Figure 2. In these same annuli the opacity and surface 
brightness were determined for each typical region. Radial annuli regardless of type of region are the dots. 
Values from annuli in the arm regions are the squares. Disk values in the inter-arm and outside regions are 
the triangles and diamonds respectively. }
\end{figure}

\end{document}